\documentclass[11pt,a4paper]{article}
\pdfoutput=1
\usepackage{myjheppub}
\usepackage{latexsym,amsfonts,amsmath,amssymb}
\usepackage{amsthm}
\usepackage{mathtools}
\usepackage{bbm}
\usepackage[abs]{overpic}
\usepackage{graphicx}
\usepackage[dvipsnames]{xcolor}
\usepackage{caption}
\usepackage{subcaption}      
\usepackage[normalem]{ulem}
\usepackage{comment}
\usepackage{url}
\usepackage{slashed}
\usepackage{cancel}
\usepackage{bm}
\usepackage{hhline}
\usepackage{hyperref}

\usepackage{tabu}

\usepackage{physics}

\usepackage[colorlinks=true
,urlcolor=blue
,anchorcolor=blue
,citecolor=blue
,filecolor=blue
,linkcolor=blue
,menucolor=blue
,linktocpage=true
]{hyperref}

\renewcommand{\Re}{\operatorname{Re}}
\renewcommand{\Im}{\operatorname{Im}}

\DeclareMathOperator{\sgn}{sgn}
\DeclareMathOperator{\Arg}{Arg}
\DeclareMathAlphabet{\mathbfsf}{OT1}{cmss}{bx}{n}

\newcommand{\Z}{\mathbb{Z}}
\newcommand{\C}{\mathbb{C}}
\newcommand{\R}{\mathbb{R}}

\newcommand{\mc}[1]{\mathcal{#1}}

\newcommand{\mf}[1]{\mathfrak{#1}}

\newcommand{\cA}{\mathcal{A}}

\newcommand{\cF}{\mathcal{F}}

\newcommand{\cH}{\mathcal{H}}
\newcommand{\cI}{\mathcal{I}}

\newcommand{\cN}{\mathcal{N}}

\newcommand{\cQ}{\mathcal Q}

\newcommand{\CN}{\mathcal{N}}
\newcommand{\CH}{\mathcal{H}}

\newcommand{\be}{\begin{equation}}
\newcommand{\ee}{\end{equation}}

\newcommand{\beq}{\begin{equation}}
\newcommand{\eeq}{\end{equation}}

\newcommand{\ii}{{\rm i}}
\newcommand{\e}{{\rm e}}

\newcommand{\rd}{{\rm d}}

\newcommand{\vol}{{\rm vol}}

\renewcommand{\j}{\varphi}

\renewcommand{\=}{\; = \;}

\allowdisplaybreaks 

\DeclareFontShape{OT1}{cmr}{mx}{n}%
{<->cmr10}{}

\DeclareMathAlphabet{\titlemath}{OT1}{cmr}{mx}{n}

\newcommand{\qE}{\mf{q}}
\newcommand{\qL}{q}
\newcommand{\aE}{\mf{a}}
\newcommand{\aL}{a}

\newcommand{\phiCanonicalPeriodicities}{\phi}
\newcommand{\BulkTheta}{\theta}
\newcommand{\BdryTheta}{\vartheta}

\newcommand{\psiCanonicalPeriodicities}{\psi}

\newcommand{\Periodictau}{t_\text{E}}
\newcommand{\IndexParameter}{\tau}

\newcommand{\EnergyOp}{E}

\newcommand{\ndt}{\noindent}

\newcommand{\upliftL}{L}

\title{The gravitational index and allowable complex metrics}

\author{Pietro Benetti Genolini${}^1$ and Sameer Murthy${}^{2}$}

\affiliation{${}^1$ D\'epartement de Physique Théorique, 
Universit\'e de Gen\`eve, 24 quai Ernest-Ansermet, \\ $\phantom{f}$ 1211 Gen\`eve, Suisse}
\affiliation{${}^2$ Department of Mathematics, King's College London, The Strand, London WC2R 2LS, UK}

\emailAdd{pietro.benettigenolini@unige.ch, sameer.murthy@kcl.ac.uk}

\abstract{
We study the Kontsevich--Segal--Witten criterion for allowable complex metrics, 
in the context of the gravitational path integral corresponding to the supersymmetric index.
In various theories of supergravity in asymptotically flat and asymptotically AdS space, 
the exponential growth of states of the corresponding microscopic index in string theory is known to 
be captured by complex saddle points of this path integral.
We compare the KSW criterion for these complex saddles 
against constraints from geometric consistency and the convergence of microscopic indices for the same saddles. 
In all the situations we consider, we find that the three criteria precisely agree with each other.
}

\begin{document}

\maketitle

\section{Introduction \label{sec:Intro}}

The Gravitational Path Integral (GPI), 
introduced in~\cite{Gibbons:1976ue},
is a very useful theoretical tool to study quantum gravity. 
The basic idea, in analogy with the path integral for quantum field theory, 
is to calculate quantum observables by summing over gravitational field configurations weighted by their classical Euclidean action. 
In the semiclassical limit the gravitational coupling is small, and the saddle points of 
the GPI correspond to solutions of the classical field equations of general relativity coupled to matter fields in the theory. 
One should then sum over saddle points including quantum fluctuations around each saddle. 

As is well-known, the perturbation theory arising from naively quantizing the metric around a saddle point is generically ill-defined, 
and it is expected that a consistent UV completion would involve new variables going beyond general relativity, as is the case in string theory. 
Nevertheless, the semiclassical GPI probes non-trivial aspects of the quantum theory in that it captures the sum 
over different geometries that is characteristic of quantum gravity. 
This simple idea can controllably predict thermodynamic transitions between different geometries, 
as in the paradigmatic example of the Hawking--Page phase 
transition between empty AdS space and AdS black holes~\cite{Hawking:1982dh}. 

However, there are many possible solutions to the equations of motion, and it is 
not always possible to ascribe a physically sensible 
interpretation to the inclusion of each saddle. 
This leads to the question of which set of saddles should be considered in the sum in first place. 
In particular, what reality properties should we assign to the metric? 
A criterion for whether a given complex metric should be included in the GPI as a physically sensible saddle point has been proposed by Witten~\cite{Witten:2021nzp},  using considerations put forward by Kontsevich--Segal~\cite{Kontsevich:2021dmb}. 
As discussed in \cite{Witten:2021nzp}, complex metrics arise in many different physical situations. 
The focus in the present article is on supersymmetric rotating black hole metrics, which generically become complex upon the analytic continuation of Lorentzian time.
Our main goal is to compare the KSW criterion~\cite{Kontsevich:2021dmb, Witten:2021nzp}
with the expectations from supersymmetric black hole entropy and the counting of their microstates in string theory.  

\medskip

\ndt {\bf The basic picture of black hole microstates in string theory}. 
The counting of supersymmetric black hole microstates, starting from the seminal works of Strominger-Vafa~\cite{Strominger:1996sh} and Sen~\cite{Sen:1995in}, is one of the big successes of string theory. 
The black holes discussed in~\cite{Strominger:1996sh, Sen:1995in} live in 
Asymptotically Flat (AF) space.  
More recently, the counting of microstates has also been understood for supersymmetric black 
holes in Asymptotically AdS (AAdS) spaces~\cite{Benini:2015eyy,Cabo-Bizet:2018ehj,Choi:2018hmj,Benini:2018ywd}.
In both types of asymptotic backgrounds, 
supersymmetric string compactifications contain a tunable parameter that 
controls the size of a black hole of given quantum numbers.
For large values of this parameter one obtains the \emph{gravitational description} of the black hole as a solution to the effective gravitational theory. 
For small values of this parameter one obtains, instead, a weakly-coupled description of microscopic degrees of freedom, which we call the \emph{microscopic description}.  
For black holes in AF space the microscopic description is given in terms of fluctuations of strings, branes, and other fundamental objects in string theory. 
For black holes in AAdS space the microscopic degrees of freedom are those of the dual CFT, as given by the AdS/CFT correspondence. 

\medskip

\ndt {\bf Supersymmetric indices with an exponential growth of states}. 
The starting point of the analysis in both AF as well as AAdS space is the calculation of a supersymmetric index in the corresponding microscopic theory. 
In AF space, the relevant indices are helicity supertraces defined in extended Poincar\'e superalgebras, see~\cite{Sen:2009vz,Dabholkar:2010rm}. 
The original Strominger--Vafa calculation has by now been extended to various situations and, in all 
cases that one can control, it is clear that the growth of states of the index agrees with 
the entropy of the supersymmetric black holes, sometimes to great accuracy~\cite{Dabholkar:2011ec,Dabholkar:2014ema,Iliesiu:2022kny},  see~\cite{Murthy:2023mbc} for a  review. 
In AAdS space the relevant indices are superconformal indices~\cite{Kinney:2005ej} or the topologically twisted index~\cite{Benini:2015noa}. 
More recent studies of these indices in different dimensions have shown that the growth of states 
also agrees with the entropy of the corresponding supersymmetric black holes~\cite{Benini:2015eyy,Cabo-Bizet:2018ehj,Choi:2018hmj,Benini:2018ywd},  see~\cite{Zaffaroni:2019dhb} for a review.

\smallskip
A crucial concept underlying these results is that of  \emph{gravitational index}, which is the supersymmetric index defined in the gravitational regime via the GPI. 
The idea is that, since the supersymmetric index is protected against changes of coupling~\cite{Witten:1982df},
we can start with the microscopic index and extrapolate it to the gravitational regime without changing its value.
The development of this topic takes two (related) routes from here. On one hand, one can zoom in to the near-horizon AdS$_2$ region of the black hole in the microcanonical ensemble, 
and show that the black hole degeneracy equals the index, thus tying up one end of the story~\cite{Dabholkar:2010rm,Iliesiu:2022kny}.\footnote{ 
Briefly, the argument relating the index to the entropy begins by showing that there is a quantum-mechanical decoupling, or energy gap, between the near-horizon AdS$_2$ region of supersymmetric extremal black holes 
and the non-supersymmetric states of the larger theory in which it is embedded (see~\cite{Mertens:2022irh} for a review). 
Then one shows that in the AdS$_2$ region the quantum theory consists only of bosonic states 
and hence the supersymmetric index equals the absolute degeneracy of states, i.e.~the exponential of the entropy~\cite{Dabholkar:2010rm,Iliesiu:2022kny}.} 
On the other hand, one can study the gravitational index in a much broader range of situations in AF and AAdS spaces in arbitrary dimensions, see~\cite{Cassani:2025sim} for a review and more details. 
It is this second context and the corresponding saddle points that form the main subject of this article.

\medskip

\ndt {\bf Complex saddles of the gravitational index}. 
The gravitational index is formally defined as the gravitational path integral 
on spaces whose asymptotic boundary contains a Euclidean thermal circle whose size corresponds to finite inverse temperature~$\beta$
around which the fermions have supersymmetric boundary conditions. 
Equivalently, we can impose boundary conditions involving complex sources for the gauge and gravitational fields 
under which the fermions are charged~\cite{Cabo-Bizet:2018ehj}.
These complex boundary conditions naturally lead to complex solutions of the field equations: 
depending on the details of the problem, either some component of the metric field or some gravitational 
charge takes complex values at the relevant saddle points. 

As we review below, examples of such saddles have been found in various settings in different dimensions and with different asymptotic conditions. 
These saddles are non-extremal, supersymmetric, complex solutions labelled by the parameter~$\beta$ corresponding to the asymptotic size of the thermal circle. 
As~$\beta \to \infty$ one recovers the Euclidean continuation of the supersymmetric extremal black hole. 
For any finite~$\beta$ the (appropriately UV-regulated) on-shell action is finite. The action is, however, independent of~$\beta$,  
consistent with the interpretation as the supersymmetric index. 
According to the canonical rules of gravitational thermodynamics~\cite{Gibbons:1976ue}, 
the on-shell action is interpreted as~$\beta$ times the grand canonical free energy. 
Although this free energy depends, in general, 
on the moduli of the theory, 
its Legendre transform agrees precisely with the microcanonical entropy of the extremal supersymmetric black hole, which is purely a function of the charges.

\medskip

\ndt {\bf The KSW criterion and the complex saddles of the index}. 
A natural question is whether there is an a priori justification for the inclusion in the GPI of these complex saddles. 
In particular, does the KSW criterion allow complex saddles that are expected to contribute to gravitational indices? Conversely, does it rule out complex solutions that are expected not to contribute? 
In this article, we compare the result of the KSW criterion applied to the complex saddles of  
gravitational indices with other physical criteria that we discuss below. 
In particular, we consider the gravitational index corresponding to 
the helicity supertrace in AF$_4$, the topologically twisted index in AAdS$_4$, and the superconformal index in~AAdS$_4$. 
The analysis of the superconformal index in~AAdS$_5$ is considered in~\cite{WIP}.
Each of these indices contains an exponential growth of states corresponding to black holes  
carrying electromagnetic charges, as well as 
angular momenta in the case of superconformal index.

The physical criteria that we impose are consistency conditions from Euclidean and Lorentzian geometry, and conditions from having convergent, well-defined microscopic indices in the boundary. 
The geometric consistency conditions include the smoothness of Euclidean sections, 
and certain conditions in the corresponding Lorentzian analytic continuation such as 
the existence of a horizon and the absence of frames rotating faster than the speed of light.
The second physical condition comes from the fact that 
the gravitational index admits a dual microscopic interpretation as a trace over a Hilbert space.
The convergence of this trace imposes additional constraints on the complex chemical potentials. 

In all cases that we consider here, we find that these conditions are 
equivalent to the KSW criterion.
In~\cite{WIP}, the same result is shown to be true for the superconformal index in AAdS$_5$. 
In other words, the three spaces cut out by geometric consistency, convergence of the trace, 
and the KSW criterion are exactly the same. 

One could compare our results with other studies of the KSW criterion with and without supersymmetry. For the complex supersymmetric black hole-like saddles in AAdS$_4$, a numerical analysis was performed in \cite{MasterThesis}, the results of which are consistent with our analysis in Section~\ref{sec:AdS4}.\footnote{We thank Davide Cassani for bringing this work to our attention.} 
In the non-supersymmetric GPI for the spectral form factor~\cite{Chen:2022hbi}, 
and for no-boundary saddles describing the origin of inflation \cite{Jonas:2022uqb, Hertog:2023vot, Maldacena:2024uhs, Janssen:2024vjn, Hertog:2024nbh},
the criterion often leads to physically sensible conditions for the inclusion of saddles. 
However, there are also solutions that violate the KSW criterion but still seem to be physically sensible, such as~\cite{Maldacena:2019cbz, Bah:2022uyz, Chen:2023hra}. 

\medskip

\ndt {\bf Brief overview of the article}. 
In Section~\ref{sec:KSW} we review the KSW criterion for the allowability of complex metrics, and its application to the metrics obtained by Wick rotation of Lorentzian rotating black holes. 
In Section~\ref{sec:Gravitational_Index}, we review the idea of the gravitational index formulated as a grand canonical partition function in gravity. 

In Section~\ref{sec:FlatSpace} we consider the supersymmetric index in AF$_4$ space, and in Section~\ref{sec:TTI} we consider the topologically twisted index in AAdS$_4$. 
In both cases we find that in, order to impose supersymmetric boundary conditions for the fermions, 
we need to perform an analytic continuation of the angular velocity and the electric potential, respectively. 
The resulting metric after Wick rotation is a gravitational instanton with real metric, and hence it trivially satisfies the KSW criterion. 
Further, we find that requiring that the convergence of the microscopic trace gives the same conditions that are imposed by smoothness of the instanton. 

In Section~\ref{sec:AdS4} we consider the superconformal index in AAdS$_4$. 
There are two potentials corresponding to the angular velocity and $U(1)_R$ potential. 
The metric of the supersymmetric saddle is complex and cannot be made real using an analytic continuation of the parameters. 
We find that the regions in parameter space carved out by requiring convergence of the trace, smoothness of the Lorentzian geometry, and the application of the KSW criterion all agree.

In Section~\ref{sec:discussion} we conclude by briefly reviewing some open questions, including the generalization to non-minimal supergravity theories,
and the relation with other allowability criteria, such as instability of branes suggested in \cite{Aharony:2021zkr}. 
For completeness, we include Appendix~\ref{app:Lorentzian_Solutions} with the Lorentzian black hole metrics in the conventions used in the paper. 

\smallskip

\noindent \textbf{Added in v2.} In a previous version of this paper, we studied the application of the KSW criterion to 
the GPI representing the superconformal index in AAdS$_5$. The results concerning AAdS$_5$ black holes in the previous version contained an error, and 
have been superseded by the results in \cite{WIP}. 
Accordingly, we have removed that section from the current version of this paper. 
We thank Oliver Janssen for communications and detailed discussions on this topic.

\section{Review of the KSW criterion}
\label{sec:KSW}

The proposal of~\cite{Witten:2021nzp} is that 
a complex metric should be \emph{allowable} if one can consistently define a generic quantum field theory on such a space, 
with the consistency condition taken to be the one earlier proposed by Kontsevich and Segal \cite{Kontsevich:2021dmb}. 
We now describe the resulting condition, which 
we refer to as KSW criterion.
Consider a smooth manifold $M$ in $d$ dimensions and the space of complex-valued metrics on it.\footnote{At 
the cost of being overly pedantic, it is worth stressing that this is \textit{not} a complexification 
of~$M$, which generically may not exist at all.} A metric~$g$ in this space is \textit{allowable} if it 
induces at each point $p$ in $M$ a complex-valued quadratic form on the real space $\Lambda^q \, T_p^*M$ such that 
\begin{equation}
\label{eq:KSW_Criterion_QuadraticForm}
    \Re \left( \sqrt{g} \, g^{i_1j_1} g^{i_2j_2} \cdots g^{i_qj_q} F_{i_1 i_2 \cdots i_q} F_{j_1 j_2 \cdots j_q} \right) \; > \; 0  
\end{equation}
for all real non-zero $q$-forms $F$, $0 \leq q \leq d$.
The condition~\eqref{eq:KSW_Criterion_QuadraticForm} can be rephrased using linear algebra~\cite{Kontsevich:2021dmb} as 
saying that at each point $p$ in $M$ one can find a basis of the real space $T_pM$ such that $g\rvert_p$ is diagonal with (a priori) complex elements $\lambda_i$ 
satisfying
\begin{equation}
\label{eq:KSW_Criterion_Eigenvalues}
    \sum_{i=1}^d \abs{\Arg \lambda_i} \; < \; \pi \, ,
\end{equation}
where $\Arg z \in (-\pi,\pi]$ is the principal value of the argument of $z$.\footnote{
The supersymmetric solutions we discuss require the existence of a spinor. To define it in general, we begin by recalling that the orthonormal frame bundle in presence of a complex metric is an $SO(d,\C)$ principal bundle, 
where $SO(d,\C)$ is the subgroup of elements of $GL(d,\C)$ preserving the complex quadratic form induced by $g$ at each point and having unit determinant. The spinor is a section of the spin bundle obtained by lifting the $SO(d,\C)$ bundle to a $Spin(d,\C)$ bundle. Here, $Spin(d,\C)$ is the complexification of $Spin(d)$, e.g. $Spin(4,\C) \cong SL(2,\C) \times SL(2,\C)$. In practice, we simply analytically continue the spinors constructed for the real metrics.}

\medskip

The spirit of the criterion is to remove negative kinetic terms (and the consequent infinite number of negative modes) in the action. 
Note that it is not strong enough to remove the existence of \emph{all} negative modes. 
A simple example illustrating this is given by the Wick-rotated Schwarzschild solution, which is a Riemannian metric on $\R^2\times S^2$ and therefore clearly satisfies the criterion, 
but suffers from the Gross--Perry--Yaffe instability due to a negative mode in the spectrum of small fluctuations around this solution~\cite{Gross:1982cv}.

\medskip

In this paper we are particularly interested in metrics that represent complex deformations of 
Wick-rotated rotating black holes. 
We begin by reviewing some examples discussed in~\cite{Witten:2021nzp}. 
The simplest examples are the line elements of four-dimensional Kerr and Kerr-AdS black holes, which can be written in the following ADM-like canonical form
\begin{equation}
\label{eq:CanonicalForm_4d_RotatingBH}
    \rd s^2 \= \beta^2 N^2 \rd \Periodictau^2 + \rho^2 \Bigl( \rd\phiCanonicalPeriodicities - \ii \beta N^{\phiCanonicalPeriodicities} \rd \Periodictau \Bigr)^2 + g_{rr} \rd r^2 + g_{\BulkTheta\BulkTheta} \rd\BulkTheta^2 \, ,
\end{equation}
in terms of the lapse function $N$ and shift vector $N^{\phiCanonicalPeriodicities}$ (as reviewed in Appendix \ref{app:Lorentzian_Solutions}).
Here $\partial_{\Periodictau}$ and $\partial_\phi$ are Killing vectors, $\Periodictau\sim \Periodictau+1$, $\BulkTheta\sim \BulkTheta + \pi$ and $\phiCanonicalPeriodicities \sim \phiCanonicalPeriodicities + 2\pi$, and $r$ is a radial coordinate.
The functions~$N, \rho, N^{\phiCanonicalPeriodicities},
g_{rr}, g_{\BulkTheta \BulkTheta}, \beta$ appearing in the metric are all real, so that in Lorentzian signature this leads to a well-defined black hole solution. 
The metric tensor~\eqref{eq:CanonicalForm_4d_RotatingBH} is complex because the shift vector is purely imaginary. 
The functions~$g_{rr}, g_{\BulkTheta\BulkTheta}, \beta$ are real and positive, and therefore 
the only part of the metric relevant to the 
application of the KSW criterion is that induced on a surface of constant $\BulkTheta$ and $r$
\begin{equation}
\label{eq:2d_Induced}
    \rd s^2 \rvert_{\rm induced} \= \Bigl( N^2 - \rho^2 \Bigl( N^{\phiCanonicalPeriodicities} \Bigr)^2 \,\Bigr) \beta^2 \rd \Periodictau^2 - 2\ii \rho^2 \beta N^{\phiCanonicalPeriodicities} \, \rd\phiCanonicalPeriodicities \, \rd\Periodictau + \rho^2  \rd\phiCanonicalPeriodicities^2 \, .
\end{equation}
One then observes that the KSW criterion \eqref{eq:KSW_Criterion_Eigenvalues} for this two-dimensional metric is equivalent to 
\begin{equation} 
\label{NNphiineq}
    - N^2 + \rho^2 \left( N^{\phiCanonicalPeriodicities} \right)^2 \; < \; 0 \, .
\end{equation}
This shows that for metrics of the type \eqref{eq:CanonicalForm_4d_RotatingBH}, the KSW criterion has a sensible physical interpretation from the Lorentzian viewpoint. 
In Lorentzian signature, \eqref{NNphiineq} is equivalent to the requirement that the norm of the Killing generator of the horizon, that is $\partial_{- \ii \Periodictau}$, is timelike everywhere outside the horizon~\cite{Witten:2021nzp}.

The inequality~\eqref{NNphiineq} does not hold for the Kerr black hole: one finds that the norm of 
the generator of the horizon has the following asymptotic behaviour, as $r\to \infty$,
\begin{equation}
    - N^2 + \rho^2 \left( N^{\phiCanonicalPeriodicities} \right)^2 \= \Omega^2 r^2 \sin^2\BulkTheta + o(r) \,.
\end{equation}
Since this expression is positive as~$r \to \infty$, there must be a surface where the sign of the norm changes and a frame corotating with the black hole cannot exist everywhere.  
The non-allowability of the Kerr metric is consistent with the instability of the thermal ensemble. Indeed, 
in the Kerr geometry, one can always have a particle (or a fluctuation of the metric) that orbits the 
black hole, say in the equatorial plane, at constant speed at a very large distance from the 
center. Such a particle carries large angular momentum compared to its energy
and, for one sign of the angular momentum, the thermal ensemble $\Tr_{\cH} \e^{ - \beta \left( H - \Omega J \right) }$ is not damped any more and therefore destabilized.

For the Kerr-AdS black hole, near the conformal boundary defined in terms of the conformal boundary coordinate $z$ as $\{ z = 0\}$, we find (see~\eqref{eq:ChangeRotatingFrame} for the change of coordinates), 
\begin{equation}
\label{eq:QuasiEuclidean_AdS_Bdry}
    - N^2 + \rho^2 (N^{\phiCanonicalPeriodicities})^2 \= - \frac{1}{z^2} \left( 1 - \Omega^2\sin^2\BdryTheta \right) + o(1) \, ,
\end{equation}
so that a frame corotating with the horizon exists all the way to the boundary, 
or, equivalently, the quasi-Euclidean metric \eqref{eq:CanonicalForm_4d_RotatingBH} is allowable, only if $\abs{\Omega} < 1$. 
Notice that this condition is equivalent to the thermodynamic stability of the thermal partition function of the dual CFT, and to the absence of superradiance~\cite{Hawking:1998kw, Hawking:1999dp}. 
Therefore, for the metrics~\eqref{eq:CanonicalForm_4d_RotatingBH}, 
the KSW criterion has a clear Lorentzian interpretation, and is also consistent with microscopic considerations. 

As we discuss in later sections, 
adding a $U(1)$ gauge field, i.e., considering the (AdS) Kerr--Newman solutions, does not change the above considerations. 
Further, the same argument goes through for higher-dimensional black holes, since it only involves the 
two-dimensional metric of the 
type~\eqref{eq:2d_Induced}.
In all these cases we obtain essentially the same  condition~\eqref{NNphiineq} 
and the subsequent conclusions~\cite{Witten:2021nzp}.
We note that rotating supersymmetric extremal black holes in AAdS$_4$ and AAdS$_5$ 
have angular velocities $\Omega = 1$, so their naive Wick rotation is not an allowed saddle of the GPI by the KSW criterion. 
Instabilities for rotating black holes in AAdS$_4$ and AAdS$_5$ have been recently revisited from 
the dual CFT point of view in~\cite{Kim:2023sig,Choi:2024xnv}.

\medskip

In the above discussion, the angular momentum is implicitly kept real.
Another possibility is to perform 
an additional analytic continuation of the parameters so that 
the shift vector becomes real: the resulting metric is a Riemannian gravitational instanton~\cite{Gibbons:1979xm} and 
the thermodynamics obtained studying this metric matches that expected of the black hole~\cite{Gibbons:1976ue}. 
This analytic continuation effectively leads to an imaginary angular momentum.
The corresponding partition function
\begin{equation}
\label{eq:Complex_PF}
    \Tr_{\cH} \, \exp \bigl( - \beta \bigl( \EnergyOp - \ii \abs{\Omega}J \bigr) \bigr) \,.
\end{equation}
has not been considered in the discussion of the 
thermal partition function in~\cite{Witten:2021nzp},
as its physical relevance in that context is not clear.

As we discuss in the remainder of the paper, partition functions with complex parameters such as~\eqref{eq:Complex_PF} do appear naturally in the context of the supersymmetric \textit{index}, 
which, as reviewed in the introduction, is a crucial observable for the counting of microstates of supersymmetric black holes. 
As mentioned above, the KSW criterion says that rotating supersymmetric extremal black holes in AAdS$_4$ and AAdS$_5$ should not be included in the gravitational path integral for the partition function.
Of course, the Wick-rotated extremal black holes have an associated infrared divergence from the infinite throat connecting the horizon to the asymptotic region. 
Regulating this divergence in a supersymmetric manner led to the complex saddles for the index in~\cite{Cabo-Bizet:2018ehj} and subsequent works. 
As we discuss below, the KSW criterion allows for these complex solutions.\footnote{We focus on the supersymmetric setting, but one may also want to consider \eqref{eq:Complex_PF} in order to investigate 
the statistics of states even outside the supersymmetric regime, e.g. setting~$\beta \abs{\Omega} = 2\pi$, see e.g.~\cite{Chen:2023mbc, Harlow:2023hjb, Benjamin:2024kdg, Grabovsky:2024vnb}.} 
This gives a different reason to consider such saddles, which may be useful in the search for saddle points which are not straightforwardly captured by the index,
such as supersymmetric grey galaxies and dual dressed black holes~\cite{Choi:2025lck}.

\section{Gravitational thermodynamics and the gravitational index}
\label{sec:Gravitational_Index}

Consider an ordinary quantum-statistical system with a Hilbert space~$\CH$, which contains 
a set of conserved charges: energy $\EnergyOp$, angular 
momenta~$J_a$, $a=1,2,\dots$, and electric charges $Q_i$, $i=1, 2, \dots$. 
In the grand canonical ensemble, we have 
chemical potentials conjugate to these charges: respectively, inverse temperature~$\beta > 0$, angular velocities~$\Omega_a$, and electric potentials~$\Phi_i$. 
The grand canonical partition function, defined as the following trace,  
\begin{equation}
\label{eq:PartitionFunction}
    Z_\text{micro}(\beta, \Omega_a, \Phi_i) \= \Tr_{\mc{H}} 
    \exp \Bigl(- \beta \EnergyOp + \beta \sum_a \Omega_a  J_a + \beta\sum_i \Phi_i \, Q_i \Bigr) \,,
\end{equation}
is an important quantity in the theory, from which many other observables can be derived. 

When there is a gravitational system dual to the above microscopic system (in either of the two senses mentioned in the introduction),
we can write the same observable as 
a path integral over gravitational field configurations with an asymptotic Euclidean time circle~$S^1$ of period~$\beta$ \cite{Gibbons:1976ue}
\begin{equation} \label{Zgravbos}
    Z_\text{grav}(\beta, \Omega_a, \Phi_i) \=  \int D g_{\mu\nu} \, D \cA^i_\mu \;  \exp \Bigl(-\int S_\text{grav}\, [g_{\mu\nu}, \cA^i_\mu] \Bigr) \,.
\end{equation}
The field content of the gravitational theory includes the metric and gauge fields for the electric charges 
(shown explicitly in~\eqref{Zgravbos}), as well as possible other fields (implicit in the notation). 
The chemical potentials~$\Phi_i$ for the electric charges are encoded in the holonomies of the gauge 
fields~$\int_{S^1} \cA_i$ at the asymptotic boundary. 
Similarly, the chemical potentials~$\Omega_a$ for angular momenta are given by the angular velocities in the gravitational theory.
The usual gravitational action for the fields implements the Hamiltonian propagation, 
and the couplings of the conserved charges to the chemical potentials are accounted for either by 
including such explicit couplings in the action 
or by twisting the charged fields of the theory around the time circle.

\vspace{0.4cm}

\ndt {\bf The gravitational supersymmetric index}

Now we turn to the supersymmetric index where, as we now explain, we have an intrinsic motivation to consider complex 
metrics. 
The simplest system in which we can discuss this is a supersymmetric 
quantum mechanics with a complex supercharge $\cQ$. 
The index is defined as a trace similar to~\eqref{eq:PartitionFunction} with the insertion of the fermion number operator~\cite{Witten:1982df}
\begin{equation}
\label{eq:defIndex}
    I_\text{micro}(\omega_b, \varphi_k) \= \Tr_{\mc{H}} \,(-1)^F  \exp \Bigl( - \beta \{\cQ,\cQ^\dagger \} +  \sum_b \omega_b \, j_b + \sum_k \varphi_k \, q_k \Bigr) \,,
\end{equation}
where the charges~$j_b, q_k$ are a subset of the charges~$J_a, Q_i$ in \eqref{eq:PartitionFunction} with 
the property that they commute with the supercharges~$\cQ, \cQ^\dagger$. 
The term in the exponent proportional to~$\beta$ is needed to define the index i.e., for the convergence of the trace. 
The fact that states in the Hilbert space that are not annihilated by~$\cQ, \cQ^\dagger$ come in boson-fermion pairs 
implies that the index~\eqref{eq:defIndex} is independent of~$\beta$ \cite{Witten:1982df}. 
Essentially the same definition of the index holds for supersymmetric quantum field theories with one complex supercharge. 

\smallskip

As mentioned in the introduction, one can express these traces as path integrals with periodic imaginary time.  
While the trace definition of the index~\eqref{eq:defIndex} is not extendable to the gravitational regime in any obvious way (as we do not know the Hilbert space), 
the path integral can at least be formally written in the gravitational variables. 
This gravitational path integral involves a spacetime with an asymptotic Euclidean time circle~$S^1$ of period~$\beta$ as in~\eqref{Zgravbos}, 
with the condition that the fermionic fields (as well as the bosonic fields, as before) have supersymmetric periodicity conditions around the circle. 
In the absence of any other twists, the fermions should be periodic to implement the~$(-1)^F$ in the trace. 

Although the GPI written initially as an integral over metrics is typically ill-defined, the supersymmetric index is 
expected to reduce to a well-defined integral over a smaller subspace of gravitational configurations that 
are annihilated by the supercharge~$\mathcal{Q}$. 
The formal arguments of localization~\cite{Witten:1988ze,Nekrasov:2002qd,Pestun:2007rz} (see the review~\cite{Pestun:2016zxk}) 
can be extended to the quantization of supergravity on non-compact spaces~\cite{deWit:2018dix,Jeon:2018kec} 
by giving an expectation value to the background superghost~\cite{Baulieu:1988xs,Costello:2016mgj,deWit:2018dix,Jeon:2018kec} 
with the result that the GPI localizes to gravitational field configurations admitting Killing spinors that agree with the fixed fields and their 
Killing spinors in the asymptotic region (i.e.~the supercharge is also allowed to fluctuate in the interior). 
We represent this integral as 
\begin{equation} \label{Igrav}
    I_\text{grav}(\omega_b, \varphi_k) \=  \int_{\scriptscriptstyle\substack{\cQ \Psi_\mu = 0 \\ \cQ \lambda = 0}} D g_{\mu\nu} \, D \Psi_\mu \, D \cA^k_\mu \, D \lambda \;  
    \exp \Bigl(-\int S_\text{grav}\, [g_{\mu\nu}, \cA^k_\mu, \Psi_\mu, \lambda] \Bigr) \,.
\end{equation}
Here we have shown the gravity, gauge fields, and their superpartners here, suppressing other possible supermultiplets in the notation.

\vskip 0.2cm

As mentioned in the introduction, the situations in which the microscopic index has an exponential growth of states are particularly interesting, as that predicts a 
black hole in the gravitational theory. 
In such situations we are faced with yet another puzzle: supersymmetric black holes
are extremal and do not contribute to the path integral with fixed~$\beta$. Relatedly, non-extremal solutions have only one 
spin structure which naively seems non-supersymmetric. 
The resolution is found by extending the potential for an R-symmetry to the imaginary plane.
Here we mean R-symmetry in the algebraic sense of any bosonic symmetry that does not commute with the global supercharge, which could be spin or global R-symmetries. 
It is clear that turning on a holonomy for an R-symmetry potential equal to~$2 \pi \ii$ 
(when the corresponding R-charge is quantized in half-integer units) 
effectively implements~$(-1)^F$ in the trace. 

In ungauged supergravity (AF spaces) the only possibility is using the angular velocity.  
In gauged supergravity (AAdS spaces), we also 
have other internal R-symmetry gauge fields 
in top-down AdS/CFT constructions in string- or M-theory, are identified with isometries of the internal space. 
See~\cite{Cassani:2025sim} for an extended discussion.
Now we have reduced the problem to a technical one: how to fill in these boundary conditions, including the 
Killing spinor boundary conditions, by smooth supersymmetric field 
configurations. In the last five years many examples of such saddles have been constructed in AAdS space~\cite{Hosseini:2017mds,Hosseini:2018dob,Cabo-Bizet:2018ehj,Choi:2018fdc,Cassani:2019mms,Kantor:2019lfo,Bobev:2019zmz,Benini:2019dyp,Nian:2019pxj,Bobev:2020pjk,Larsen:2021wnu,BenettiGenolini:2023rkq,BenettiGenolini:2023ucp}
as well as AF space~\cite{Iliesiu:2021are,Hristov:2022pmo,H:2023qko,Boruch:2023gfn,Cassani:2024kjn,Boruch:2025qdq}.

In each case, the resulting metric is a supersymmetric but non-extremal solution depending on~$\beta$. 
However, the gravitational action of this metric is independent of~$\beta$ and reproduces the known saddle point value of the expected microscopic index. 
In fact, in the context of gauged four-dimensional supergravity with vector multiplets, 
one can calculate the action using equivariant localization 
even in the absence of explicit analytic expressions for the solution \cite{BenettiGenolini:2023kxp}, 
and show that it is independent of~$\beta$ when the spacetime topology includes a cigar factor ~\cite{BenettiGenolini:2024hyd, BenettiGenolini:2024lbj}. 
A generic feature of these saddles is that the field configurations are complex in some way. 
This should not be surprising: since we have given a chemical potential a complex value, the charge at the saddle point generically has a complex value as well. 
This leads us to naturally consider reality conditions that are different from what one usually 
imposes, and this is a good place to test different allowability criteria.

\section{Supersymmetric index in \texorpdfstring{AF$_4$}{AF4} space}
\label{sec:FlatSpace}

One of the simplest supersymmetry-protected observable one can compute using semiclassical gravity
are supersymmetric indices in theories with~$\CN = 2$ 
supersymmetry in four-dimensional flat space. 
Such indices are realized concretely in compactifications of Type II string theory on a Calabi--Yau threefold.

One begins by considering a complex supercharge in such theories which satisfies 
\begin{equation}
\label{eq:KN_BPSBound}
    \{ \cQ, \cQ^\dagger \} \= \EnergyOp - \EnergyOp_\text{BPS} \,, 
\end{equation}
where~$\EnergyOp$ is the energy, and~$\EnergyOp_\text{BPS}$ is the 
so-called BPS energy, given by the central charge of the theory, 
which, in general, is a function of the electric and magnetic charges as well as the moduli of the theory.
The above expression takes positive value on generic (long) multiplets of the superalgebra, and vanishes 
for $\frac12$-BPS (short) multiplets. 
Short multiplets have four states that are related by fermion zero modes. 

The simplest index that gets contributions only from short multiplets is called the second helicity supertrace, 
defined as a Witten index with two insertions of the spacetime helicity. 
The insertions of the helicity operator effectively absorb the fermion zero modes 
to give a non-zero answer for short multiplets~\cite{Bachas:1996bp, Gregori:1997hi}. 
After absorbing the fermion zero modes, the index in the canonical ensemble with fixed electric charge~$Q$ and inverse temperature~$\beta$ is defined as the following trace over the Hilbert space~$\cH_Q$\footnote{One can also introduce a fixed magnetic charge~$P$ though, for simplicity, we set it to zero here.}
\begin{equation}
\label{eq:GravitationalIndex}
    \cI(\beta, Q) \= \Tr_{\cH_{Q}} (-1)^F \e^{ - \beta \{ \cQ, \cQ^\dagger\} } \e^{-\beta \EnergyOp_\text{BPS}(Q)} \,.
\end{equation}
Similar helicity supertraces capture BPS states in~$\CN \ge 2$ superalgebras. 
These indices have been explicitly calculated in compactifications of string theory with 
$\CN=4$ and~$\CN=8$ supersymmetry, 
using a weakly-coupled description in which the Hilbert space is known, see the review~\cite{Mandal:2010cj}.

The gravitational description of these theories is given in terms of low energy supergravity coupled to matter. 
The theory relevant for generic black hole solutions is~$\CN=2$ ungauged supergravity coupled to vector multiplets, whose field content consists of the metric, gauge fields, scalars, and their superpartners. 
The index~\eqref{eq:GravitationalIndex} can be computed in the gravitational theory using a partition function in a mixed ensemble depending on~$\beta$, the angular velocity~$\Omega$, and fixed electromagnetic charges. 
We consider solutions of four-dimensional $\cN=2$ ungauged supergravity (potentially with additional matter) 
that preserve supercharges $\cQ$, $\cQ^\dagger$ obeying the algebra~\eqref{eq:KN_BPSBound}. 
We require the solutions to be asymptotically flat with the following falloff: outside a compact region we impose that the underlying manifold is $\R_+ \times S^1 \times S^2$ with $\R_+$ parametrized by $r$ and, as $r\to \infty$, the metric is asymptotically
\begin{equation}
\label{eq:KN_BoundaryConditions_1}
    \rd s^2 \sim \rd r^2 + \beta^2 \rd\Periodictau^2 + r^2 \Bigl( \rd\BulkTheta^2 + \sin^2\BulkTheta \bigl( \rd\phiCanonicalPeriodicities - \ii \beta \Omega \, \rd\Periodictau \bigr)^2 \Bigr) \, ,
\end{equation}
where $\Periodictau \sim \Periodictau + 1$, $\BulkTheta \sim \BulkTheta + \pi$, $\phiCanonicalPeriodicities \sim \phiCanonicalPeriodicities + 2\pi$, and we impose
\begin{equation}
\label{eq:KN_BoundaryConditions_2}
    \beta \Omega = 2\pi\ii (1+2n) \, , \quad n \in \Z \, .
\end{equation}
This condition, which is at the origin of the $(-1)^F$ in \eqref{eq:GravitationalIndex}, guarantees that the spinors are anti-periodic around the $S^1$ parametrized by $\Periodictau$.
Moreover, we require that the graviphoton gauge field has electric charge $Q_e$ through $S^2$ as $r\to \infty$.
It is straightforward to check that the partition function in this ensemble coincides with \eqref{eq:GravitationalIndex}, i.e., 
\begin{equation}
\label{eq:GravitationalIndex_2}
\begin{split}
    \Tr_{\cH_{Q_e}} \exp \bigl( - \beta \EnergyOp + \beta\Omega J \bigr) 
    &\= \Tr_{\cH_{Q}} (-1)^{2J} \exp \bigl( - \beta \{ \cQ, \cQ^\dagger\} - \beta \EnergyOp_\text{BPS} \bigr) \\
    &\= \cI(\beta, \EnergyOp_\text{BPS}) \, .
\end{split}
\end{equation}

As is well-known, extremal $\frac12$-BPS black hole solutions of~$\CN=2$ ungauged supergravity are spherically symmetric and are described in terms of the attractor mechanism~\cite{Ferrara:1995ih}. 
The attractor mechanism shows that, near the horizon of the black hole, the scalar fields gain a mass, and 
the effective description is given in terms of the graviton and the single graviphoton multiplet. 
At two-derivative level, the action is governed by the action of minimal supergravity. 
These conclusions rely quite crucially on the extremality and spherical symmetry of the supersymmetric black hole solutions. 

As mentioned above, our focus here is on non-extremal solutions that are saddle points to the gravitational index. 
In a bit of surprise, it was shown  in~\cite{Boruch:2023gfn, Boruch:2025qdq} that 
these index saddles also obey a form of the attractor mechanism, dubbed the \emph{new attractor mechanism}.
Although the generic solutions depend on the moduli and the temperature, and break spherical symmetry by a rotation, 
the moduli fields at the fixed points of the rotation on the horizon are fixed in terms of the charges of 
the solution, and the contribution of the saddles to the index are also locally independent of the moduli. 
As in the extremal case, all these features can be mapped to the simple case of supersymmetric solutions in minimal ungauged supergravity, exactly as in the classic attractor mechanism.  
Following these ideas, we focus on supersymmetric  solutions in the simplest theory, minimal ungauged supergravity in the following presentation.

\medskip

The bosonic fields of minimal~$\CN=2$ supergravity
are the metric and a $U(1)$ gauge field $\cA$ with curvature $\cF=\rd\cA$ interacting via the bosonic action\footnote{We set $G_N = 1$ in the following.}
\begin{equation}
\label{eq:Action_Ungauged}
    S \= - \frac{1}{16\pi} \int \left( R - \cF^2 \right) \, \vol \, .
\end{equation}
The central charge is given by the electric charge~$Q_e$, and the relevant superalgebra and the index that we study are given by~\eqref{eq:KN_BPSBound} and~\eqref{eq:GravitationalIndex}, respectively. 
A supersymmetric solution supports a Dirac spinor $\epsilon$ satisfying the equation\footnote{Here and in all the supersymmetric solutions that we discuss in this paper,
we consider the analytic continuation of the Killing spinors that solve the Killing spinor equation in Lorentzian signature. In particular, there could be
more general solutions to the Killing spinor equation in Euclidean signature, see~\cite{Boruch:2023gfn} for a more detailed discussion of this setup.}
\begin{equation}
\label{eq:KSE_Ungauged}
    \left( \nabla_\mu + \frac{\ii}{4} \cF_{\nu\rho}\gamma^{\nu\rho}\gamma_\mu \right) \epsilon \= 0 \, ,
\end{equation}
where $\gamma_\mu$ generate Cliff$(4,0)$.

The saddle point solutions to the index~\cite{Boruch:2023gfn,Boruch:2025qdq} are
supersymmetric solutions that belong to the family of Israel--Wilson--Perjes metrics \cite{Israel:1972vx,Perjes:1971gv,Hartle:1972ya,Tod:1983pm, Whitt:1984wk}, 
and can be expressed in the following form,  
\begin{align}
\label{eq:KN_SUSY_Metric}
\begin{split}
\rd s^2 &\= \frac{\Delta_r}{B} \beta^2 \rd \Periodictau^2 + W \biggl( \frac{\rd r^2}{\Delta_r} + \rd \BulkTheta^2  \biggr) \\
& \quad  + \sin^2\BulkTheta \, B \biggl( \rd\phiCanonicalPeriodicities + \aE \beta \frac{ \Delta_r \left( r_+^2 - \aE^2 \cos^2\BulkTheta \right) + (r^2 - \aE^2) (r^2 - r_+^2)  }{(r_+^2 - \aE^2) B W} \rd \Periodictau \biggr)^2  \, ,
\end{split} \\[10pt]
\label{eq:KN_SUSY_GaugeField}
\cA &\= - \ii \frac{ \qL r }{W}\Bigl( \beta \bigl( 1 - \ii \aE \sin^2\BulkTheta \, \Omega \bigr) \, \rd \Periodictau + \aE \sin^2\BulkTheta \,  \rd\phiCanonicalPeriodicities \Bigr) + \ii \beta \Phi_e \, \rd \Periodictau  \, .
\end{align}
The functions appearing here are given by 
\begin{equation}
\label{eq:KN_SUSY_MetricFunctions}
    \Delta_r  \= (r - \qL)^2 - \aE^2 \, , \qquad 
    W \= r^2 - \aE^2 \cos^2 \BulkTheta \, , \qquad  
    B  \;\equiv \; \frac{(r^2 - \aE^2)^2 + \aE^2 \sin^2\BulkTheta \, \Delta_r}{W} \, , 
\end{equation}
the chemical potentials are  
\begin{equation}
     \Omega \= \frac{\ii\aE}{r_+^2 - \aE^2} \, , \qquad \Phi_e \=  \frac{r_+ \qL }{r_+^2 - \aE^2} \, , 
\end{equation}
and inverse temperature is given by 
\begin{equation}
    \beta \= 4\pi \frac{r_+^2 - \aE^2}{\Delta'_r(r_+)} 
    \= \pm 2 \pi \frac{r_+^2 - \aE^2 }{\aE} \, .
\end{equation}

The parameter~$r_+$ is a solution to $\Delta_r = 0$, which is easily solved to give 
\begin{equation}
\label{eq:KN_rPlus}
    r_+ \= \qL \pm \aE \, .
\end{equation}
The $\pm$ sign labels the two branches of solutions and appears in the expression for $\beta$ as well. 
The metric has been written in the canonical form~\eqref{eq:CanonicalForm_4d_RotatingBH}, highlighting that~\hbox{$N^{\phiCanonicalPeriodicities}\rvert_{r=r_+}=0$}. This family of supersymmetric solutions depends on two parameters, $\qL$ and $\aE$.
The metric tensor is real provided $\qL$ and $\aE$ are real, and it is clear that if $\BulkTheta \sim \BulkTheta + \pi$ and $\phiCanonicalPeriodicities \sim \phiCanonicalPeriodicities+2\pi$ are spherical coordinates on $S^2$, $\Periodictau\sim \Periodictau + 1$, and $r$ is real and positive, then these solutions match the boundary conditions \eqref{eq:KN_BoundaryConditions_1} and \eqref{eq:KN_BoundaryConditions_2} with $n=0,-1$. It is also straightforward to compute the ADM mass and electric charge of the solutions and check that $\EnergyOp=Q_e$, matching the BPS bound \eqref{eq:KN_BPSBound}.\footnote{Here and throughout, the conserved charges and chemical potentials of the complex solutions are defined via analytic continuation of those of the real Lorentzian solutions reviewed in appendix \ref{app:Lorentzian_Solutions}.} Therefore, they are good candidates to be semiclassical saddles of the GPI corresponding to the required microscopic description.

Additionally, in order for the metric tensor to be defined smoothly on the $\R^2 \times S^2$ manifold, 
we need $r \geq r_+$, and $r_+$ to be the largest root of $\Delta_r$, that is, 
on the ``positive'' branch (choice of upper sign in \eqref{eq:KN_rPlus}) we need $\mf{a}>0$, 
and on the ``negative'' branch (choice of lower sign in \eqref{eq:KN_rPlus}), we need $\mf{a}<0$.

It is convenient, in order to compare with the asymptotically AdS case studied in the following sections, 
to exchange the parameters $(\qL, \aE)$ for $(r_+, r_\star)$. The parameter~$r_\star$ is the extremal radius, 
which is the value of $r_+$ for which $\beta$ diverges, which then requires $\aE=0$. 
That is, the parameters are defined by \eqref{eq:KN_rPlus} and
\begin{equation}
\label{eq:KN_rStar}
    r_\star \;\equiv \; q \, ,
\end{equation}
so $\aE = \pm ( r_+ - r_\star)$.
In terms of these parameters, we can write
\begin{equation}
\label{eq:KN_SUSY_Quantities}
\begin{aligned}
    \beta &\= 2\pi \frac{r_\star (2 r_+ - r_\star )}{ r_+ - r_\star } \, , &\qquad
    \Omega &\= \pm \frac{ \ii(r_+ - r_\star) }{ r_\star ( 2 r_+ - r_\star ) } \, , &\qquad
    \Phi_e &\= \frac{r_+ }{ 2 r_+ - r_\star } \, .
\end{aligned}
\end{equation}

\medskip

The solutions \eqref{eq:KN_SUSY_Metric}, \eqref{eq:KN_SUSY_GaugeField} can be obtained from the Lorentzian Kerr--Newman spacetime parametrized by $(m,a,q)$ as follows \cite{Yuille:1987vw} 
(for completeness, the reader can find this solution with our conventions in Appendix~\ref{app:Lorentzian_Solutions}). 
First, we perform a Wick rotation, obtaining the complex metric discussed in Section~\ref{sec:KSW}, which does not satisfy the KSW criterion. 
We then impose supersymmetry: integrability of \eqref{eq:KSE_Ungauged} requires $m = \qL$. 
This also means that the ADM mass and the electric charge of the solution are related by $\EnergyOp = Q_e$~\cite{Gibbons:1982fy}. 
Importantly, in order to have a spinor defined on the disc $(r,\Periodictau)$, we need it to be anti-periodic as $\Periodictau \sim \Periodictau + 1$. As discussed earlier, this needs
\begin{equation}
\label{eq:AF_Constraint}
    \beta \, \Omega \; \in \; 2\pi \ii \, (1+2n) \, , \quad n \;\in \; \Z \, .
\end{equation}
This condition does not fix $a$.
However, requiring that the radial coordinate is real and positive, and so is $r_+$, means that $a$ should be analytically continued to $a=\ii \aE$, 
and correspondingly leads to a pure imaginary angular velocity (and therefore the angular momentum is also pure imaginary).
This leads to a Riemannian metric, as originally considered by Gibbons and Hawking \cite{Gibbons:1979xm}, and thus trivially satisfies the KSW criterion. 
Note, however, that in this case supersymmetry necessary leads to a pure imaginary, non-zero value of angular velocity, even at asymptotic infinity.

Some additional properties of this family of solutions are worth mentioning, as they will also apply to the following cases. 
First, taking the analytic continuation of \eqref{eq:KN_SUSY_Metric} using $\beta\Periodictau = \ii t$ and $\aE = - \ii a$ leads to a real Lorentzian supersymmetric metric that does not describe a black hole but a naked singularity.  
On the other hand, one can take the extremal limit of \eqref{eq:KN_SUSY_Metric}, corresponding to $r_+ \to r_\star$ (or, equivalently, $\aE=0$). 
Upon taking the analytic continuation of the resulting solution again using $\beta\Periodictau=\ii t$, one does obtain the regular supersymmetric extremal Reissner--Nordstr\"om solution (see Appendix \ref{app:Lorentzian_Solutions}).
This is consistent with the fact that supersymmetric Lorentzian black holes are extremal (see for instance \cite{Cvetic:2005zi} for a review in gauged supergravity). 

\medskip

We conclude this section with a remark about geometric constraints. We have used $(r_+, r_\star)$ to parametrize the ensemble, but the gravitational index \eqref{eq:GravitationalIndex} is defined in a mixed ensemble parametrized by~$(\beta, Q_e)$. Inverting the relations we find
\begin{equation}
    r_\star \= Q_e \, , \qquad r_+ \= Q_e \frac{ \beta - 2 \pi Q_e }{\beta - 4 \pi Q_e } \, .
\end{equation}
As discussed earlier, it is well-motivated from the geometry (namely, by requiring that the metric tensor is defined on $\R^2\times S^2$) to choose $r_+$ to be the largest positive root of $\Delta_r$, or, equivalently, to choose a specific sign for $\aE$ on the two branches, which is also equivalent to impose
\begin{equation}
\label{eq:4d_AF_SUSY_Constraint}
    r_+  \, > \, r_\star \, > \, 0 \, .
\end{equation}
Note that, in terms of $Q_e$ and $\beta$, these conditions are equivalent to 
\begin{equation}
    Q_e \,> \, 0 \, , \qquad \beta - 4\pi Q_e \,>\, 0 \, .
\end{equation}
This inequality was also discussed in the context of slightly different geometric constraints in~\cite{Iliesiu:2021are, Boruch:2023gfn}.

\section{Topologically twisted index in \texorpdfstring{AAdS$_4$}{AAdS4} space} 
\label{sec:TTI}

We now move to four-dimensional asymptotically (locally) anti-de Sitter spacetime, for which the 
microscopic construction of the gravitational supersymmetric index is provided by the dual SCFT$_3$. 
In this section we consider
the topologically twisted index of this SCFT$_3$~\cite{Benini:2015noa, Benini:2016hjo, Closset:2016arn}.

\medskip

Accordingly, we consider a three-dimensional $\cN=2$ theory on~$S^1\times \Sigma_g$, where~$S^1$ has circumference~$\beta$ and~$\Sigma_g$ is a Riemann surface of genus~$g >1$ 
with the constant curvature metric.\footnote{There is also a topologically twisted index on $S^2$, which admits a refinement weighing the states by the value of their angular momentum on the sphere. The bulk contribution would be given by rotating dyonic supersymmetric solutions with $S^2$ horizon. However, there are no known such solutions in minimal gauged supergravity. Solutions of this family are explicitly known in the STU model, but only with vanishing temperature \cite{Hristov:2018spe}. The on-shell action of the supersymmetry-preserving non-extremal deformation can be computed indirectly \cite{BenettiGenolini:2024hyd, BenettiGenolini:2024lbj}.}
The theory generically has a~$U(1)_R$ R-symmetry, and a global flavor symmetry group~$G_F$ 
whose Cartan Lie algebra is generated by~$J_F^\alpha$,  $\alpha=1, \dots, {\rm rk}\, {\rm Lie}(G_F)$.
The supersymmetry-preserving background is constructed by turning on an R-symmetry background gauge field $A_R$ 
with flux through $\Sigma_g$,  
\begin{equation}
\label{eq:TTI_RSymmetry_Flux}
	\frac{1}{2\pi}\int_{\Sigma_g} \rd A_R \= g-1 \, ,
\end{equation}
which implements the topological twist~\cite{Benini:2015noa, Benini:2016hjo, Closset:2016arn}. 
We impose that the background gauge field also has non-trivial holonomy around the circle $S^1$, 
\begin{equation}
\label{eq:TTI_RSymmetry_Holonomy}
	-\frac{1}{2\pi} \int_{S^1} A_R \; \equiv \; \frac{\beta\Phi_R}{2\pi\ii} \= \frac{1+2n}{2} \, , \qquad n \in \Z \, .
\end{equation}
More generally, the holonomy around the circle takes value in~$\Z_2$ and encodes the choice of spin structure on the circle \cite{Closset:2018ghr}. 
Both choices are consistent with supersymmetry, and here we choose this holonomy to be non-trivial. 
This imposes that the fermions in the system are anti-periodic around~$S^1$, which is more natural from the bulk point of view, as we see below. 

The quantization of the theory on~$\Sigma_g$ leads to a Hilbert space~$\cH_{\Sigma_g}$ whose states are labelled by their energy~$\EnergyOp$ and 
of the~$J_F^\alpha$, and the integer $R$-charges (as follows from the twist condition \eqref{eq:TTI_RSymmetry_Flux}). The 
supercharges preserved by the background obey the anti-commutation relation
\begin{equation}
\label{eq:SUSY_TTI}
    \{ \cQ, \cQ^\dagger \} \= \EnergyOp - 2\pi \sum_\alpha \sigma^\alpha_F J^\alpha_F \, ,
\end{equation}
where $\sigma_\alpha$ are the real masses in the background vector multiplets for the flavor symmetries. 
The topologically twisted index in the holographic dual SCFT$_3$ is defined as the thermal partition function on this background, and has the form 
\begin{equation}
\label{eq:TTI_Defn}
\begin{split}
	&\Tr_{\cH_{\Sigma_g}} \exp \Bigl( - \beta \EnergyOp + \beta \Phi_R R + \beta \sum_\alpha \Phi_F^\alpha J_F^\alpha \Bigr) \\
	&\qquad \= \Tr_{\cH_{\Sigma_g}} (-1)^{R} \exp \Bigl( - \beta \{ \cQ, \cQ^\dagger \} + 2\pi\ii \sum_\alpha u_F^\alpha J_F^\alpha \Bigr) \\
    &\qquad \; \equiv \; \cI_{\rm TT}(u^\alpha_F) \, , 
\end{split}
\end{equation}
where $u^\alpha_F \equiv \beta (\Phi_F^\alpha - 2\pi \sigma^\alpha_F)/2\pi\ii$.
Notice that the presence of a grading by the $R$-charge of the states, due to the non-trivial holonomy \eqref{eq:TTI_RSymmetry_Holonomy}. 
As stated above, this is equivalent to the spinors in the system being anti-periodic around $S^1$. 
This corresponds to the only smooth spin structure on the circle that can bound a disc, 
and thus allow a gravity dual with the topology of a black hole~\cite{BenettiGenolini:2023ucp}.
This will also appear when discussing the superconformal index.
For simplicity, in the following we will not consider the refinement by flavor symmetry.

In order to construct the three-dimensional background on $S^1\times \Sigma_g$, we begin with the geometry $S^1\times H^2$ and a non-trivial R-symmetry background gauge field given by 
\begin{equation}
\label{eq:TTI_Bdry_Background}
\begin{split}
    \rd s^2 \= \beta^2 \rd \Periodictau^2 + \rd\theta^2 + \sinh^2\theta \, \rd \phi^2 \, , \qquad A_R \= \ii \beta \Phi_R \, \rd \Periodictau + \frac{1}{2} \cosh\theta \, \rd\phi \, . 
\end{split}
\end{equation}
We then quotient this hyperbolic geometry by discrete subgroups of $SO(1,2)$, in order to obtain the Riemann surface~$\Sigma_g$. 
Note that the constraint~\eqref{eq:TTI_RSymmetry_Flux} is implemented by this quotient construction. 
This metric and gauge field background is real if and only if $\beta$ is real and~$\Phi_R$ is pure imaginary.
The second condition follows from the first due to the constraint~\eqref{eq:TTI_RSymmetry_Holonomy}.

\medskip

To look for supersymmetric bulk contributions to the GPI with the boundary conditions~\eqref{eq:TTI_Bdry_Background} we consider 
minimal gauged supergravity.
The bosonic fields are the metric and a $U(1)$ gauge field $\cA$ as in the 
previous section, but now the action also includes a negative cosmological constant equal to $-3/\ell^2$,
\begin{equation}
\label{eq:MinimalGaugedSUGRA_Action}
    S \= - \frac{1}{16 \pi} \int \Bigl( R + \frac{6}{\ell^2} - \cF^2 \Bigr) \vol \, .
\end{equation}
Supersymmetry of the solution requires the existence of a Dirac spinor $\epsilon$ satisfying
\begin{equation}
\label{eq:MinimalGaugedSUGRA_KSE}
    \Bigl( \nabla_\mu - \frac{\ii}{\ell} \cA_\mu + \frac{1}{2\ell} \gamma_\mu + \frac{\ii}{4} \cF_{\nu\rho} \gamma^{\nu\rho} \gamma_\mu \Bigr)\, \epsilon = 0 \, .
\end{equation}
In the following, we set $\ell=1$. It can be reinstated using dimensional analysis from the formulae in Appendix \ref{app:Lorentzian_Solutions}.

The relevant supersymmetric saddles we discuss here belong to the following 
family, labelled by the real charge~$\qE$, 
\begin{align}
\label{eq:TTI_Bulk_SUSY_Soln_Metric}
    \rd s^2 &\= \beta^2 V(r) \, \rd \Periodictau^2 + \frac{\rd r^2}{V(r)} + r^2 \left( \rd\theta^2 + \sinh^2\theta \, \rd \phi^2 \right) \, , \qquad V(r) \= \left( r - \frac{1}{2r} \right)^2 - \frac{\qE^2}{r^2} \, , \\
\label{eq:TTI_Bulk_SUSY_Soln_GaugeField}
    \cA &\= \beta\left( \frac{\qE}{r} + \ii\Phi_e \right) \rd \Periodictau + \frac{1}{2} \, \cosh\theta \, \rd\phi \, , \qquad \cF \= \beta\frac{\qE}{r^2} \, \rd \Periodictau \wedge \rd r + \frac{1}{2} \, \sinh\theta \, \rd\theta \wedge \rd\phi \,.
\end{align}
Here, as above, we assume a quotient of the $H^2$ in order to compactify the horizon, and 
the electric potential and the inverse temperature are given by 
\begin{equation}
\label{eq:TTI_Phi}
    \Phi_e \= \frac{\ii\qE}{r_+} \, , 
    \qquad \beta \= \frac{4\pi}{V'(r_+)} \= \pm \pi \frac{r_+}{\qE} \, . 
\end{equation}
The parameter $r_+$ satisfies $V(r_+)=0$, and we can use this to exchange $\qE$ for $r_+$, obtaining two branches of solutions
\begin{equation}
\label{eq:TTI_rh}
    \qE \= \pm \left( r_+^2 - \frac{1}{2} \right) \, ,
\end{equation}
where the sign is the same appearing in \eqref{eq:TTI_Phi}.
These solutions are supersymmetric for all values of $r_+$ and it is clear that 
if we identify $\Periodictau \sim \Periodictau + 1$, the conformal boundary conditions $r\to \infty$ match~\eqref{eq:TTI_Bdry_Background} upon identifying $\Phi_R = \Phi_e$. 
Moreover, the solutions also satisfy the constraint~\eqref{eq:TTI_RSymmetry_Holonomy}, since $\beta \Phi_e = \pm \pi \ii $. 
It is straightforward to use canonical methods of holographic renormalization to compute the conserved charges of this solution, 
finding that the mass of the solution vanishes. 
This matches the BPS relation obtained from \eqref{eq:SUSY_TTI}~\cite{Hristov:2011ye}. 
Therefore, they are good candidates to be semiclassical saddles of the GPI representing~\eqref{eq:TTI_Defn}. 

Moreover, these solutions are real and regular with topology $\R^2 \times \Sigma_g$ provided $r>r_+>1/\sqrt{2}$. As before, we can define $r_\star$ to be the value of $r_+$ such that $\beta$ diverges, namely $r_\star = 1/\sqrt{2}$. In terms of these parameters, we find
\begin{equation}
\label{eq:TTI_ChemPot}
    \Phi_e \= \pm \frac{\ii}{\sqrt{2} r_+ r_\star} ( r_+^2 - r_\star^2) \, , \qquad \beta \= 2\pi \frac{r_+ r_\star^2}{r_+^2 - r_\star^2} \, .
\end{equation}
It is clear that, under the assumptions above,
\begin{equation}
\label{eq:TTI_PositiveBeta}
    \beta \, > \, 0 \quad \Longleftrightarrow \quad r_+ \, >\, r_\star \, ,
\end{equation}
which, as just pointed out, is the condition needed for regularity of the metric as well.
The relevance of this will be clear presently.

\medskip

The solution \eqref{eq:TTI_Bulk_SUSY_Soln_Metric}, \eqref{eq:TTI_Bulk_SUSY_Soln_GaugeField} can be obtained from Lorentzian static dyonic black holes with horizon $\Sigma_g$ (see \cite{BenettiGenolini:2019jdz, Bobev:2020pjk, BenettiGenolini:2023ucp} and Appendix \ref{app:Lorentzian_Solutions} for more details on the solutions). 
The Lorentzian solution is labelled by the three real parameters $(\eta, \qL, p)$, corresponding to the energy and electric and magnetic charges. One first performs a Wick rotation $t=-\ii\beta\Periodictau$, obtaining a complex metric. 
Requiring supersymmetry imposes two conditions via the integrability of \eqref{eq:MinimalGaugedSUGRA_KSE}: $\eta = 0$, and $p^2 = 1/4$ \cite{Caldarelli:1998hg}, thus fixing the energy of the solution to vanish, and fixing the magnetic charge in terms of the topology. 
The global existence on the $\R^2$ factor of the spinor imposes that the latter is anti-periodic as $\Periodictau \sim \Periodictau + 1$. In this gauge, this imposes
\begin{equation}
\label{eq:Filippo}
    \frac{\beta\Phi_e}{2\pi\ii} \; \in \; \frac{1+2n}{2} \,, \quad n \in \Z \, .
\end{equation}
which is satisfied with $n=0,-1$ for any $\qL$. 
Here, as for rotating black holes, we require that the radial coordinate is real and positive, and that so is $r_+$, which imposes the analytic continuation $\qL=\ii \qE$, making both the resulting metric and gauge field real for both branches of \eqref{eq:TTI_rh}.\footnote{We could also allow for $r_+$ and~$\beta$ to be complex. 
In this case, we can study the application of the KSW criterion to this metric allowing for the coordinate $r$ to trace a path in complex plane, 
starting from complex $r_+$ and asymptotically becoming real, as done for (AdS-)Schwarzschild black holes in \cite{Chen:2022hbi}. 
The result is that along this specific contour the KSW criterion becomes~$\Re\beta > 0$.
}
Therefore, the KSW criterion is trivially satisfied by these saddle points of the gravitational path integral, although this choice is quite non-trivial from the Lorentzian viewpoint. Indeed, while the analytic continuation of \eqref{eq:TTI_Bulk_SUSY_Soln_Metric} via $\beta \Periodictau=\ii t$ and $\qE = - \ii \qL$ is a real Lorentzian metric, it does not generically describe a black hole, but a static dyonic singularity. It is only in the extremal limit $\qL\to 0$ that $V(r)$ admits a real solution and one recovers the regular supersymmetric extremal black hole with $\Sigma_g$ horizon.\footnote{As in the previous section, we point out that the same metric and gauge field can be obtained by taking the extremal limit directly in \eqref{eq:TTI_Bulk_SUSY_Soln_Metric}, \eqref{eq:TTI_Bulk_SUSY_Soln_GaugeField} and only then taking the Wick rotation $\beta \Periodictau=\ii t$.}

It is also important to notice that the condition~\eqref{eq:TTI_PositiveBeta} is naturally 
imposed by the microscopic definition of the index~\eqref{eq:TTI_Defn}.
Indeed, in absence of flavor refinement, the convergence of the trace that guarantees the possibility of restricting to the BPS subsector of states requires $\Re\beta > 0$.
Since~$\beta$ at the saddle point~\eqref{eq:TTI_Phi} is real under our assumptions, this is equivalent to the condition~\eqref{eq:TTI_PositiveBeta}.

\section{Superconformal index in  \texorpdfstring{AAdS$_4$}{AAdS4} space}
\label{sec:AdS4}

In this section we discuss the holographic gravitational calculation of the three-dimensional superconformal index. 
The framework for the discussion in this section is related to the previous section in that we also have asymptotically AdS$_4$ spacetime.
However, there are differences between the two discussions that we comment on in the following presentation.

\medskip

We consider a three-dimensional~$\cN=2$ SCFT on~$S^1\times S^2$, 
where the~$S^1$ has circumference~$\beta$ and the~$S^2$ has unit radius. 
The generator of the azimuthal~$U(1)\subset SO(3)$ is denoted by~$J$. The theory has a~$U(1)_R$ R-symmetry whose generator is denoted~$R$, and a global flavor symmetry 
group~$G_F$ whose Cartan generators are denoted by~$J_F^\alpha$, $\alpha=1,2,\dots, \text{rk}(G_F)$. 
The following background for the metric and R-symmetry gauge field preserves two supercharges, 
\begin{equation}
\label{eq:3d_SCI_Background}
    \rd s^2 \= \beta^2 \rd \Periodictau^2 + \rd\BdryTheta^2 + \sin^2\BdryTheta \bigl( \rd\phiCanonicalPeriodicities - \ii \beta\Omega \, \rd\Periodictau \bigr)^2 \, , \qquad A_R \= \ii \beta\Phi_R \, \rd\Periodictau \,.
\end{equation}
Here the coordinates are identified as~$\Periodictau \sim \Periodictau + 1$, $\BdryTheta \sim \BdryTheta + \pi$, $\phiCanonicalPeriodicities \sim \phiCanonicalPeriodicities + 2\pi$.
In contrast to the case discussed in the previous section, here the R-symmetry background gauge field 
has vanishing flux through~$S^2$, and therefore the two backgrounds are topologically distinct~\cite{Closset:2013vra}. 
It is also possible to turn on flat background gauge fields for the flavor symmetry $A^\alpha_F = \ii \beta \Phi^\alpha_F \, \rd\Periodictau$, but we do not consider this here.

Quantization of the theory on $S^2$ leads to a Hilbert space $\cH_{S^2}$ with states labelled 
by~$\EnergyOp,J,J^\alpha_F,R$, which are, respectively,
the eigenvalues of the Hamiltonian, angular momentum, the flavor symmetry, and the R-symmetry. 
If we only consider abelian R-symmetry, the corresponding charge does not have to be quantized, in contrast to 
Section~\ref{sec:TTI}.
The two supercharges have the following anti-commutation relation, 
\begin{equation}
\label{eq:3d_SCI_BPS_Bound}
    \{ \cQ, \cQ^\dagger\} \= \EnergyOp - J - \frac{1}{2}R \, .
\end{equation}
The spinors are anti-periodic around $S^1$ if the parameters of the background \eqref{eq:3d_SCI_Background} satisfy
\begin{equation}
\label{eq:3d_SCI_Antiperiodicity}
    \frac{\beta}{2\pi\ii } \bigl( 1-4\Phi_R + \Omega \bigr) \= 1 + 2 n \, , \qquad n \in \Z \, .
\end{equation}

On this background we consider the partition function 
\begin{equation}
\label{eq:3d_SCI_Definition}
\begin{split}
   & \Tr_{\cH_{S^2}} \exp \Bigl( - \beta \EnergyOp + \beta \Omega J + \beta \Phi_R R + \sum_\alpha \beta \Phi_F^\alpha J_F^\alpha \Bigr) \\
    &\qquad \= \Tr_{\cH_{S^2}} (-1)^{2J} \exp \Bigl( - \beta \{ \cQ, \cQ^\dagger \}  + \beta \left( 4 \Phi_R - 2 \right) \Bigl( J + \frac{1}{4}R \Bigr) + \sum_\alpha \beta \Phi_F^\alpha J_F^\alpha \Bigr) \\
    &\qquad \= \cI_{\rm SC} \biggl( \frac{\beta}{2\pi\ii }(4\Phi_R - 2) \mp 1, \frac{\beta}{2\pi\ii} \Phi^\alpha_F \, \biggr) \, ,
\end{split}
\end{equation}
which, as indicated in the last line, corresponds to the superconformal index, defined as 
\begin{equation}
\label{eq:3d_SCI_Defn}
    \cI_{\rm SC}(\IndexParameter, \varphi_F^\alpha) \; \equiv \;  \Tr_{\cH_{S^2}} (-1)^{\frac{R}{2}} \exp \Bigl( - \beta \{ \cQ, \cQ^\dagger \}  + 2\pi\ii \IndexParameter \Bigl( J + \frac{1}{4} R \Bigr) + 2\pi\ii \sum_\alpha \j_F^\alpha J_F^\alpha  \Bigr) \, .
\end{equation}
Here we have used an $R$-graded definition of the superconformal index, in contrast to the superconformal index graded by $2J$ defined in \cite{Bhattacharya:2008zy,Bhattacharya:2008bja}. The different grading is reflected in a different weight of the states in the trace and, 
as discussed in \cite{Choi:2019dfu, GonzalezLezcano:2022hcf, Bobev:2022wem, BenettiGenolini:2023rkq}, the two indices are related by a shift of $\IndexParameter$ by~$\pm 1$. 
The same weighting applied also to the topologically twisted index in \eqref{eq:TTI_Defn}, though there was no angular momentum in that case.
The superconformal index is a more complicated object than the topologically twisted index~\eqref{eq:TTI_Defn} 
and, in order to see the microstate degeneracy,
we need to take a ``Cardy-like'' limit by tuning~$\IndexParameter$ as well as a large-$N$ limit~\cite{Choi:2019zpz,BenettiGenolini:2023rkq}. 
A careful analysis of the asymptotic behaviour of the index shows that the $R$-graded 
index~\eqref{eq:3d_SCI_Defn} has a leading contribution at $\tau\to 0$.\footnote{The ``generalized Cardy limit'' 
with $\IndexParameter\to d/c\in\mathbb{Q}$ generically corresponds to sub-leading contributions. For ABJM theory, 
the gravitational dual saddles involve orbifolding the internal $S^7$~\cite{BenettiGenolini:2023rkq}.}

We remark that the supersymmetry-preserving background~\eqref{eq:3d_SCI_Background} is inherently complex if~$\Omega \neq 0$, in contrast with the three-dimensional 
background of the topologically twisted index~\eqref{eq:TTI_Bdry_Background} and the solutions~\eqref{eq:KN_SUSY_Metric} relevant for the gravitational index. This is because one cannot choose~$\Omega$ and~$\Phi_R$ pure imaginary, as this 
would be inconsistent with the constraint \eqref{eq:3d_SCI_Antiperiodicity}.\footnote{One 
could also calculate the same index~$\mathcal{I}_\text{SC}$ using a untwisted background, i.e.~with~$\Omega=0$ 
as in the approach of~\cite{Closset:2013vra}. 
Note, however, that even in that approach, the Killing vector constructed as a bilinear 
in the supercharges is intrinsically complex, and it generates two independent isometries of~$S^1\times S^2$.
The Killing vector cannot be made real without changing the transversely holomorphic foliation and hence the 
supersymmetric background structure~\cite{Closset:2019hyt}.
It is then reasonable to conjecture that the two backgrounds are related by a supersymmetry-preserving 
deformation~\cite{ArabiArdehali:2021nsx,  Cassani:2021fyv, BenettiGenolini:2023rkq}.
A step towards proving this has been taken in~\cite{Inglese:2023tyc} by studying complex 
supersymmetry-preserving backgrounds.\label{footnote:Qdeformations_3d}}
Because of the complex nature of the boundary, we expect to find complex gravitational fillings. 

\medskip

As in the previous case, we consider the gravitational dual to the superconformal index without flavor 
refinements, thus restricting our attention to the minimal gauged supergravity 
theory~\eqref{eq:MinimalGaugedSUGRA_Action}.
The relevant supersymmetric solutions we find belong to the following family
\begin{align}
\label{eq:AdSKN_SUSY_Metric}
\begin{split}
\rd s^2 &\= \frac{\Delta_r\Delta_\BulkTheta}{B\, \Xi^2} \beta^2  \rd \Periodictau^2 + W \left( \frac{\rd r^2}{\Delta_r} + \frac{\rd \BulkTheta^2}{\Delta_\BulkTheta}  \right) \\
& \ \ \  + \sin^2\BulkTheta \, B \Bigg( \rd\phiCanonicalPeriodicities - \ii a \frac{ \Delta_r \left( r_+^2 + a^2\cos^2\BulkTheta \right) + \Delta_{\BulkTheta} (r^2 + a^2) (r^2-r_+^2)  }{ (r^2_+ + a^2) B W \Xi} \beta \rd \Periodictau \Bigg)^2  \, ,
\end{split} \\[10pt]
\label{eq:AdSKN_SUSY_GaugeField}
\cA &\= \frac{ m r \sinh \delta}{W\,\Xi}\Big( -\ii \beta \left( \Delta_\BulkTheta - a \sin^2\BulkTheta \, \Omega \right) \, \rd \Periodictau - a \sin^2\BulkTheta \,  \rd\phiCanonicalPeriodicities \Big) + \ii \beta \Phi_e \, \rd \Periodictau  \, .
\end{align}
The quantities in the metric and gauge field are given by
\begin{equation}
\label{eq:AdSKN_SUSY_MetricFunctions}
\begin{split}
    \Delta_r &\= (r^2 + a^2)(1 + r^2) - 2 m r \cosh \delta + m^2 \sinh^2 \delta \, , \\
    \Delta_\BulkTheta &\= 1 -  a^2 \cos^2 \BulkTheta \, , \qquad W \= r^2 + a^2 \cos^2 \BulkTheta \, , \qquad \Xi\= 1 - a^2 \, , \\
    B &\;\equiv \; \frac{\Delta_\BulkTheta(r^2+a^2)^2 - a^2 \sin^2\BulkTheta \, \Delta_r}{W \Xi^2} \, , \\
    \beta &\= 4\pi \frac{a^2 + r_+^2}{\Delta'_r(r_+)} \= - 2\pi r_+ \frac{a^2 + r_+^2}{ r_+^2 (1 + a^2) - 3r_+ m \cosh\delta + 2m^2 \sinh^2\delta } \, , \\    
    \Omega &\= a\frac{1 + r_+^2}{a^2 + r_+^2} \, , \qquad \Phi_e \=  \, \frac{m r_+ \sinh \delta }{a^2 + r_+^2} \, .
\end{split}
\end{equation}
In contrast to the cases considered in Section \ref{sec:FlatSpace} and \ref{sec:TTI}, where the metric tensor was real and we could use Riemannian geometry, here the metric tensor is complex. 
It is defined on the manifold~$\R^2\times S^2$, parameterized by~$(r,\Periodictau)$ and~$(\BulkTheta, \phiCanonicalPeriodicities)$ respectively, 
where~$\BulkTheta \sim \BulkTheta + \pi$ and~$\phiCanonicalPeriodicities \sim \phiCanonicalPeriodicities+2\pi$ 
are spherical coordinates on~$S^2$, 
and~$r>0$ and $\Periodictau \sim \Periodictau + 1$ are the polar coordinates on~$\mathbb{R}^2$. 
The quantity $r_+$ is a positive root of $\Delta_r=0$, as we discuss below. 
Note that the metric has been written in the canonical form~\eqref{eq:CanonicalForm_4d_RotatingBH}, 
which requires~$N^{\phiCanonicalPeriodicities}\rvert_{r=r_+}=0$ so that~$\phiCanonicalPeriodicities$ can be a real coordinate.

The parameters in the solution are constrained by \cite{Kostelecky:1995ei, Caldarelli:1998hg}
\begin{equation}
\label{eq:AdSKN_SUSY_BPSBound}
    a \= \coth\delta -1 \, .
\end{equation}
Thus, the above family of supersymmetric solutions is labelled by two parameters $(m,\delta)$. 
It is more convenient to exchange them for the two parameters $(r_+, r_\star)$, where $r_+$ is defined as above, 
and $r_\star$ labels the location of the horizon of the supersymmetric extremal black 
hole.\footnote{\label{ft:notradius4d} Note that there is no spherical symmetry, so $r_\star$ is not a ``radius'' of the horizon. Indeed, the entropy of the black hole has the form
\begin{equation}
    S = \frac{\pi}{G_4} \frac{r_\star^2}{1-r_\star^2} \, ,
\end{equation}
which isn't bounded even if $r_\star$ is.}
The precise relations are
\begin{align}
    \label{eq:AdSKN_SUSY_rStar}
    \coth\delta &\= r_\star^2 +1 \, , \\
    \label{eq:AdSKN_SUSY_m}
    m \sinh\delta &\= r_+ \Bigl(1 + r_\star^2 \, \Bigr) \pm \ii \abs{ r_+^2 - r_\star^2 } \, .
\end{align}
The $\pm$ sign above indicate the two branches of solutions of the quadratic equation for $m$ coming from  $\Delta_r(r_+)=0$. 
In obtaining \eqref{eq:AdSKN_SUSY_m} as a solution, we have assumed that~$r_+$ and~$r_\star$ are real and positive.
We continue with these assumptions in the remaining part of the analysis. 
Note that, even in the previous sections, we implicitly assumed that the radial coordinate was real.
In terms of $r_\star$ and $r_+$, we find that the potentials have the form
\begin{equation}
\label{eq:4d_AdSKN_ThermodynamicPotentials_SUSY}
\begin{aligned}
\beta &\=  2\pi \frac{ r_+^2 + r_\star^4 }{(r_+^2 - r_\star^2) ( (1 + r_\star^2)^2 + 4 r_+^2 ) } \left[ 2 r_+ \pm \ii \left( 1 + r_\star^2 \right) \sgn (r_+ - r_\star)  \right] \, , \\
\Omega &\= r_\star^2 \frac{1+ r_+^2}{r_+^2 + r_\star^4} \, , \\
\Phi_e &\= \frac{r_+^2 (1 + r_\star^2) \pm \ii \abs{ r_+^2 - r_\star^2 } r_+}{r_+^2 + r_\star^4 } \,,
\end{aligned}
\end{equation}
where again the labels $\pm$ refer to the two branches.

In order to match the boundary conditions described earlier, we should take $r\to \infty$, and it is convenient to introduce the following coordinate change \cite{Hawking:1998kw}
\begin{equation}
    \label{eq:ChangeRotatingFrame}
    \begin{aligned}
    \frac{\cos\BdryTheta}{z} &\= r \cos\BulkTheta \, , &\qquad \frac{1}{z^2} &\= \frac{r^2\Delta_\BulkTheta + a^2\sin^2\BulkTheta}{\Xi} \, .
    \end{aligned}
\end{equation}
The metric and gauge field \eqref{eq:AdSKN_SUSY_Metric} and \eqref{eq:AdSKN_SUSY_GaugeField} then have the following behavior at leading order
\begin{equation}
\label{eq:AdS4asymp_v0}
\begin{split}
\rd s^2 & \; \sim \; \frac{\rd z^2}{z^2} + \frac{1}{z^2} \left[ \beta^2 \rd \Periodictau^2 + \rd\BdryTheta^2 + \sin^2\BdryTheta \bigl( \rd\phiCanonicalPeriodicities - \ii \beta \Omega \, \rd \Periodictau \bigr)^2 \right] \, , \\
    A & \; \sim \; \ii \beta \Phi_e \, \rd \Periodictau \, .
\end{split}
\end{equation}
Moreover, it is straightforward to check that the potentials \eqref{eq:4d_AdSKN_ThermodynamicPotentials_SUSY} satisfy
\begin{equation}
\label{eq:4d_AdSKN_SUSYConstraint}
    \frac{\beta}{2\pi\ii} \left( 1 - 2\Phi_e + \Omega \right) \= \mp 1 \, ,
\end{equation}
and that the conserved charges computed using holographic renormalization satisfy
\begin{equation}
\label{eq:4d_AdSKN_SUSYAlgebra}
    \EnergyOp - J - Q_e \= 0 \, .
\end{equation}
Thus, this family of solutions matches the SCFT BPS bound~\eqref{eq:3d_SCI_BPS_Bound} and the boundary conditions~\eqref{eq:3d_SCI_Background} and~\eqref{eq:3d_SCI_Antiperiodicity} with $\Phi_e = 2\Phi_R$. 
Moreover, this allows us to identify in terms of gravitational quantities the variable of the superconformal index~\eqref{eq:3d_SCI_Background}. Upon comparing with~\eqref{eq:3d_SCI_Definition}, we find
\begin{equation}
\label{eq:3d_SCI_Parameter}
    \IndexParameter \= \frac{\beta}{2\pi\ii } (2\Phi_e - 2) \mp 1 \= \frac{\beta}{2\pi\ii} \left( \Omega - 1 \right) \, .
\end{equation}
Therefore, we notice that the connection between the field theory variables and the gravity potentials goes through ``reduced'' potentials, namely
\begin{equation}
\label{eq:4d_AdSKN_ReducedChemicalPotentials}
\begin{split}
    \IndexParameter_g &\; \equiv \;\frac{\beta}{2\pi\ii} ( \Omega - \Omega_\star) \= \frac{ \mp \sgn(r_+ - r_\star) \left( 1 - r_\star^4 \right) +  2\ii \left( 1 - r_\star^2 \right) r_+ }{ \left( 1 + r_\star^2 \right)^2 + 4 r_+^2 }  \, , \\
    \varphi_g & \; \equiv \; \frac{\beta}{2\pi\ii} ( \Phi_e - \Phi_{e\star}) \\
    & \= \frac{\pm \left[ 1 + 2 (r_\star^2 + 2r_+^2) + r_\star^4 - \sgn(r_+ - r_\star) ( 1 - r_\star^4 ) \right] + 2\ii r_+ \left( 1 - r_\star^2 \right)}{ 2 \bigl( \left( 1 + r_\star^2 \right)^2 + 4 r_+^2 \bigr) } \, ,
\end{split}
\end{equation}
where~$\Omega_\star = 1$ and $\Phi_{e\star} = 1$ are the values of the angular velocity and electrostatic potential for the extremal solution. 
We see from~\eqref{eq:3d_SCI_Parameter} that $\IndexParameter = \IndexParameter_g$ and the supersymmetry constraint~\eqref{eq:4d_AdSKN_SUSYConstraint} 
becomes~$\IndexParameter_g - 2\varphi_g = \mp 1$, confirming again that the contribution to the 
gravitational path integral of this family of solutions is a function of a single complex variable.

\medskip

The solutions described by \eqref{eq:AdSKN_SUSY_Metric} and~\eqref{eq:AdSKN_SUSY_GaugeField} 
can be obtained from the Lorentzian AdS-Kerr--Newman solution by Wick rotation~$t=-\ii\beta \Periodictau$ and imposing~\eqref{eq:AdSKN_SUSY_BPSBound} 
(see Appendix~\ref{app:Lorentzian_Solutions}).\footnote{We note that the Lorentzian metric of the Kerr--Newman black hole can be obtained from the Lorentzian metric of the AdS-Kerr--Newman black hole in the limit that $\ell$ is much larger than any other length scale in the solution (namely $m$, $a$ and $q$). 
However, taking the analogous limit in the supersymmetric solution described in this section requires $r_+/\ell \to 0$ and $r_\star/\ell \to 0$, which necessarily takes us to the extremal limit of the solutions described in section \ref{sec:FlatSpace} (that is, $r_+ \to r_\star$). 
For instance, it's straightforward to see from~\eqref{eq:4d_AdSKN_ThermodynamicPotentials_SUSY} that $\beta \to \infty$ and $\Omega \to 0$.} As mentioned, the resulting spacetime satisfy the supersymmetry algebra \eqref{eq:4d_AdSKN_SUSYAlgebra} and have a globally well-defined spinor, as guaranteed by \eqref{eq:4d_AdSKN_SUSYConstraint}. 
However, upon Wick-rotating back to ``Lorentzian'' signature with $\beta \Periodictau = \ii t$, the resulting solution is complex and does not describe a causally well-behaved black hole, 
unless one also requires that the solution is extremal, in which case the metric becomes real and Lorentzian. This solution, which is the same obtained via Wick-rotation of the extremal limit of \eqref{eq:AdSKN_SUSY_Metric}, is the supersymmetric extremal rotating electrically charged black hole with spherical horizon at~$r=r_\star$ \cite{Kostelecky:1995ei, Caldarelli:1998hg}. 
Because of this interpretation, in keeping with the historical case of the Kerr spacetime~\cite{Gibbons:1976ue}, 
we assume that the coordinate~$r$ is real, and that $r_+$ and $r_\star$ are both real and positive. 
These complex supersymmetric saddles of the GPI defined by the superconformal index have been studied in~\cite{Cassani:2019mms, Bobev:2019zmz, Benini:2019dyp, Nian:2019pxj, BenettiGenolini:2023rkq}. 

\medskip

In contrast to the cases considered in the previous sections, the semiclassical saddles of the GPI dual to the superconformal index~\eqref{eq:3d_SCI_Definition} are complex, 
and so constitute a good testing ground for the KSW criterion. Our main interest is the relation between the constraints in the parameter space $(r_\star, r_+) \in \R_{\geq 0} \times \R_{\geq 0}$ imposed by three different viewpoints on the superconformal index. 
\begin{enumerate}
\item The first viewpoint is the microscopic definition~\eqref{eq:3d_SCI_Defn}. Since the 
eigenvalues of $\{ \cQ, \cQ^\dagger\}$ are non-negative and in principle unbounded above, 
a well-defined trace over the full Hilbert space requires~$\Re \beta >0$.  
Once we have defined it properly, we note the well-known fact that the trace is actually independent of~$\beta$ and collapses to 
a sum over the BPS states which are annihilated by~$\cQ$ and~$\cQ^\dagger$~\cite{Witten:1982df}. 
Since, generically, operators can carry arbitrarily large spin and R-charge, we have that~$J + \frac{1}{4}R$ is unbounded from above. We assume that it is bounded below on the BPS subspace, this assumption holds in all examples that we have studied (see e.g.~\cite{BenettiGenolini:2023rkq}). 
Now the convergence of the trace on the BPS subspace imposes that~$\Im \tau >0$.  
On the specific gravitational saddle, $\beta$ is given by the expression~\eqref{eq:4d_AdSKN_ThermodynamicPotentials_SUSY} and~$\tau$ is given by the expression for~$\tau_g$ in~\eqref{eq:4d_AdSKN_ReducedChemicalPotentials}. 
We have
\begin{equation}
\label{eq:4d_AdSKN_Constraints_Micro}
    \Im \tau_g >0 \; \Longleftrightarrow \; r_\star < 1 \, , \qquad \Re \beta >0  \; \Longleftrightarrow \; r_\star < r_+  \, .
\end{equation}

\item The second viewpoint is the allowability criterion~\eqref{eq:KSW_Criterion_Eigenvalues} on the above sub-family of solutions.
The result of our analysis presented below is that the KSW criterion also carves out the same region~\eqref{eq:4d_AdSKN_Constraints_Micro}. 
\item The third viewpoint is the analytic continuation to~\eqref{eq:AdSKN_SUSY_Metric} of the geometric properties of the Lorentzian black hole solutions.
The AdS-Kerr--Newman solutions are well-behaved black holes only if the following conditions are satisfied~\cite{Hawking:1998kw}. 
Firstly, we should have~$a^2 < 1$, since otherwise~$\Delta_{\BulkTheta}$ and hence the 
metric becomes degenerate at some point.
On the supersymmetric locus, it is immediate 
to verify that this translates into~
\begin{equation}
\label{eq:4d_AdSKN_rstar_Constraint}
    r_\star \,< \, 1 \, .
\end{equation}
Secondly, we require the absence of velocity of light surfaces, 
where the Killing generator of the horizon becomes null outside the horizon, which means that $ \abs{\Omega}<1$.
On the supersymmetric locus, including the condition~\eqref{eq:4d_AdSKN_rstar_Constraint},  
this is equivalent to 
\begin{equation}
\label{eq:AdSKN_VelocityLight_Condition_SUSY}
    r_+  > r_\star \,, \qquad  r_\star < 1 \,
\end{equation}
Note that $r_+ > r_\star$ is also naturally imposed from the smoothness of the geometry of the complex saddle, 
namely it ensures that $r_+$ is the largest positive root of $\Delta_r$.\footnote{We considered the same constraint also in the asymptotically flat case \eqref{eq:4d_AF_SUSY_Constraint} and in the topologically twisted index in AdS \eqref{eq:TTI_PositiveBeta} (and we also implicitly assumed it in e.g.~\cite{BenettiGenolini:2023rkq}).}

\end{enumerate}

It is notable that the three conditions (i) well-definedness (convergence) of the microscopic partition function, (ii) KSW criterion for the complex saddles, and (iii) definition of the Lorentzian metric and absence of velocity of light surfaces lead to the same conditions \eqref{eq:AdSKN_VelocityLight_Condition_SUSY} on the parameters.
This conclusion is analogous to what was found for the quasi-Euclidean examples discussed in~\cite{Witten:2021nzp} 
and reviewed in Section~\ref{sec:KSW}: 
in those examples, the KSW criterion, requiring the absence of velocity of light surfaces, and requiring that the thermal trace is well-defined all gave 
the same condition, namely $\abs{\Omega}<1$.
Here the actual conditions are slightly more involved, since we have a complex metric (rather than having only a pure imaginary shift vector), 
and in particular $\beta$ is generically complex, but the agreement between the three criteria remains. 
This is because $\Omega$ in \eqref{eq:4d_AdSKN_ThermodynamicPotentials_SUSY} is real, and therefore there is a straightforward relation between $\Re\beta$ and $\Im\IndexParameter$ , namely
\begin{equation}
    \Im \IndexParameter_g \= \frac{1 - \Omega_g}{2\pi} \Re\beta_g \, .
\end{equation}

\medskip

In the remainder of the section, we show how the KSW criterion can be applied to the supersymmetric complex saddles. It is a daunting task to obtain explicit expressions for analytic application of the 
criterion to the metric on the entire spacetime, so we present below a combination of analytic and numerical results.

In order to apply the KSW criterion, we notice that assuming that $r_+$ and $r_\star$ are real means that $a=r_\star^2$ is real, and therefore~$g_{\BulkTheta\BulkTheta}$ and~$\Omega$ are also real.
Since there are no off-diagonal terms in~$\BulkTheta$, it is enough to consider the metric induced on a surface of constant~$\BulkTheta$, so that the problem is effectively three-dimensional. 

\bigskip

\ndt {\bf Analysis near the horizon}

\smallskip

We begin the analysis
by considering the region near the locus \hbox{$\{ r = r_+\}$}.
Expanding in powers of $R^2 = r-r_+$, 
we find, to leading order, 
\begin{equation}
\label{eq:AdSKN_4d_NHMetric}
\begin{split}
    \rd s^2_3 &\; \sim \; \frac{4 W(r_+)}{\Delta_r'(r_+)}\, \Bigl(\rd R^2 + (2\pi)^2 R^2 \, \rd\Periodictau^2  \Bigr) + \sin^2\BulkTheta \frac{\Delta_\BulkTheta (r_+^2 + r_\star^4)^2}{W(r_+) \; \Xi^2} \rd\phiCanonicalPeriodicities^2 \, .
\end{split}
\end{equation}
With our assumptions, the coefficient of~$\rd\phiCanonicalPeriodicities^2$
is real and so we only need to look at the first term, which is conformal to flat space. 
Since the radial coordinate~$r$ is taken to be real, 
the allowability 
criterion~\eqref{eq:KSW_Criterion_Eigenvalues} reduces to
\begin{equation} 
\label{eq:KSWRsmall}
    \pi \, > \, 2 \abs{\Arg ( \Delta'_{r}(r_+))} \,.
\end{equation}
For the two branches of solutions introduced below~\eqref{eq:AdSKN_SUSY_rStar} 
we have  
\begin{equation}
    \Delta'_r(r_+) \=  (r^2_+ - r_\star^2) \left[ 4 r_+ \mp 2\ii \sgn (r_+ - r_\star) ( 1+r_\star^2) \right] \, .
\end{equation}
Assuming that $r_+$ and $r_\star$ are positive, we find that the KSW criterion \eqref{eq:KSWRsmall} is equivalent to requiring that the real part of~$\Delta'_r(r_+)$ is positive, or $r_+ > r_\star$, which is equivalent to one of the two conditions defining the region \eqref{eq:AdSKN_VelocityLight_Condition_SUSY}.
Furthermore, from the definition of~$\beta$ in~\eqref{eq:AdSKN_SUSY_MetricFunctions} and~\eqref{eq:4d_AdSKN_ThermodynamicPotentials_SUSY}, we notice that \eqref{eq:KSWRsmall} is also equivalent to requiring that the real part of $\beta$ is positive. 
In particular, the limiting case is that of pure imaginary $\beta$. This is a particularly relevant case for the analysis of Kontsevich--Segal \cite{Kontsevich:2021dmb}, as the Lorentzian metric would belong to the boundary of the allowable region and would be causally well-behaved.

\bigskip

\ndt {\bf Analysis in the asymptotic region}

\smallskip

In the asymptotic region, as $r\to \infty$, it is convenient to use the coordinates \eqref{eq:ChangeRotatingFrame}, such that the metric takes the form \eqref{eq:AdS4asymp_v0}, namely
\begin{equation} 
\label{eq:AdS4asymp}
\begin{split}
    \rd s^2_3 & \; \sim \; \frac{\rd z^2 }{z^2} + \frac{1}{z^2} \Bigl( \beta^2 \left( 1 - \Omega^2 \sin^2\BdryTheta \right) \rd \Periodictau^2 + \sin^2\BdryTheta \, \rd\phiCanonicalPeriodicities^2 - 2\ii \beta \Omega \sin^2\BdryTheta \, \rd \phiCanonicalPeriodicities \, \rd \Periodictau \Bigr) \, .
\end{split}
\end{equation}
The~$g_{zz}$ component of the metric is clearly real, so we are left with an effective two-dimensional problem in the~$\phiCanonicalPeriodicities$--$\Periodictau$ plane.

In this case, it is easier to consider the criterion in the form \eqref{eq:KSW_Criterion_QuadraticForm} for the corresponding two-dimensional metric $g_{(2)}$, keeping $\sin\BdryTheta>0$, that is, outside the degenerate points. The $q=0$ case imposes that $\Re \sqrt{\beta^2\sin^2\BdryTheta}>0$, and since we have already found from the analysis of the metric near the horizon that $\Re\beta >0$, this instructs us to take the square root with the~$+$ sign. The $q=1$ case imposes that the matrix $\Re W \equiv \Re \sqrt{g_2} \, g_{(2)}^{-1}$ is positive definite. Using the fact that  $\Omega\in\R$ 
(see~\eqref{eq:AdSKN_SUSY_MetricFunctions}), we obtain the diagonal matrix
\begin{equation}
\begin{split}
    \Re W &\= \begin{pmatrix}
        \sin\BdryTheta\Re \frac{1}{\beta} & 0 \\ 0 & \frac{1}{\sin\BdryTheta} \left( 1 - \Omega^2 \sin^2\BdryTheta \right) \Re \beta 
    \end{pmatrix} \, .
\end{split}
\end{equation}
Since~$\Re \, \beta > 0$ and $\Re \, \beta^{-1} >0$ are equivalent conditions, 
the condition~$\Omega^2<1$ implies that~$W$ is positive definite. 
Thus, we find that the KS criterion applied to the metric near the conformal boundary is equivalent to the conditions~\eqref{eq:AdSKN_VelocityLight_Condition_SUSY}.

\bigskip

\ndt {\bf Analysis in the interior region}

\smallskip

As mentioned above, at a generic point in the bulk we have to consider the eigenvalues of a three-dimensional metric on the space parametrized by the real coordinates $(\Periodictau, r, \phi)$. 
We have checked numerically that the KSW criterion~\eqref{eq:KSW_Criterion_Eigenvalues} is satisfied 
if and only if the constraints~\eqref{eq:AdSKN_VelocityLight_Condition_SUSY} hold.
More precisely, we selected~$10^6$ random values for~$(r_\star,r_+,r,\BulkTheta)$ subject to the conditions $0 < r_\star < 1$, $ r_\star < r_+ < r < 10^3$, 
and $\BulkTheta\in (0,\pi)$,  
and verified  that~\eqref{eq:KSW_Criterion_QuadraticForm} holds
by numerically evaluating both the $q=0$ and $q=1$ criteria. 
These results are consistent with those obtained in~\cite{MasterThesis}.

\section{Discussion and open questions \label{sec:discussion}}

The KSW criterion cuts out an allowed region in the space of parameters of gravitational 
partition functions containing black holes~\cite{Witten:2021nzp}. 
The allowed potentials are precisely those corresponding to thermodynamic stability of the partition function in the grand canonical ensemble. 
For example, in AdS space the condition on the angular velocity is~$|\Omega|<1$, which is precisely the condition of convergence of the thermal trace. 
The corresponding instability can be traced to modes of large angular momenta arising from particles rotating around the black hole very far from the horizon. 

In this paper we have studied the analogous phenomena for supersymmetric black holes. 
The gravitational partition functions containing these black holes are also not convergent---in the 
above example in AdS space they have~$|\Omega|=1$, thus lying just outside the allowed region.  
However, the path integral for the gravitational index has additional imaginary potentials turned on at infinity. 
The convergence of the thermal-type trace translates to the positivity of imaginary parts 
of the potentials dual to charges and angular momenta. 

We find that the KSW criterion is equivalent to the convergence of the microscopic 
trace (and also to geometric criteria) in all examples.

\medskip

There are many interesting questions  that could be investigated, even within the range of supersymmetric black objects. 
One broad point is that we could test ideas about quantum gravity and the swampland. 
For example, various microscopic indices are known to contain black holes and 
black strings in compactifications of M-theory, and F-theory on Calabi--Yau manifolds~\cite{Maldacena:1996gb,Haghighat:2015ega}. 
Taking these as data points for consistent quantum gravitational calculations, one could test the KSW criterion against them.
On the other hand, complex saddles have been shown to play a role in describing different 
black objects like black strings, branes~\cite{Chen:2024gmc,Cassani:2024kjn, Boruch:2025qdq}, and 
spindles~\cite{Cassani:2021dwa} in supergravity. It would be interesting to see what 
the KSW criterion says about these low-energy calculations. 

Complex solutions have appeared in the context of Euclidean supergravity also outside the context of supersymmetric indices (for instance, they naturally appear as bulk duals to field theories on spheres in presence of mass deformations \cite{Freedman:2013oja, Bobev:2013cja}). 
Their contribution to the relevant observables (e.g.~the renormalized free energy) matches the result obtained from supersymmetric localization on the field theory side, so we expect that they would satisfy the KSW criterion. 

The grand canonical partition functions we consider are defined as a sum over gravitational saddles with appropriate boundary conditions required by supersymmetry. As remarked in~\cite{Aharony:2021zkr, Iliesiu:2021are, Boruch:2022tno}, these are invariant under integer shifts: for instance, the condition~\eqref{eq:3d_SCI_Antiperiodicity} is invariant under
\begin{equation}
\label{eq:ABJM_Shifts}
      \Phi_R \, \to \, \Phi_R + \frac{2\pi\ii}{\beta} n_R \, , \quad \Omega \, \to \, \Omega + \frac{2\pi\ii}{\beta} n_{\Omega} \, , \quad \text{provided } \ 4n_R + n_\Omega \in 2\Z \, . 
\end{equation}
Therefore, a priori, we should also include an infinite number of saddle points in addition to the supersymmetric solutions discussed in this paper. However, including them leads to inconsistencies, as observed in \cite{BenettiGenolini:2023rkq}, extending to four dimensions the arguments in \cite{Aharony:2021zkr}. The authors of \cite{Aharony:2021zkr}, who studied the case of AAdS$_5 \times S^5$ black holes in Type IIB supergravity, pointed out that it is possible to restrict the shifted contributions by considering the effect of branes wrapped on Euclidean cycles in the ten-dimensional geometry. Such a brane-stability criterion is currently unknown for the AAdS$_4\times S^7$ black holes in M-theory, but if it exists it would be worthwhile studying its relation to the KSW criterion.

\medskip

There are also natural generalizations of the discussions in this paper that could be addressed. 
We comment on some of them below. 
\begin{itemize}
    \item In our discussion of the supersymmetric indices presented in the paper, we did not include refinements constructed using global flavor symmetry groups. 
    This brought us to consider only minimal supergravities describing only the interaction of the graviton multiplet. 
    It is possible to include additional vector multiplets corresponding to flavor refinements of the dual indices, and some solutions are explicitly known. 
    For the topologically twisted index discussed in Section~\ref{sec:TTI}, one can include a $U(1)$ refinement, 
    for which the bulk gravity dual is the four-dimensional $X^0X^1$ model, and there are solutions described in \cite{BenettiGenolini:2023ucp}, which are supersymmetric non-extremal deformations of dyonic black holes with two electric charges. Supersymmetry fixes the magnetic charge in terms of the AdS radius, and the electric charges are equal. 
    For these, it is straightforward to see that the conclusion is the same as that obtained in Section~\ref{sec:TTI}. 
    They become real Euclidean solutions with topology $\R^2 \times \Sigma_g$ upon performing the analytic continuation 
    of the electric charge that is required by the supersymmetry condition, so again supersymmetry imposes the allowability. 
    More interesting and technically more involved are the supersymmetric non-extremal deformations of electric rotating black holes in the $X^0X^1$ model 
    that are described in~\cite{Cassani:2019mms, BenettiGenolini:2023rkq}. 
    They are dual to the $U(1)$ refinement of the superconformal index described in Section~\ref{sec:AdS4}. 
    It would be interesting to apply the KSW criterion to those solutions and investigate whether it persists the relation described 
    in Section~\ref{sec:AdS4} between convergence of the partition function and allowability. 
    \item We discussed the role of the KSW criterion in the selection of the metric in the saddle points of the GPI. 
    One could also wonder about the role played by the Abelian gauge field that appears in all our solutions. 
    For the AF$_4$ supersymmetric saddles discussed in Section \ref{sec:FlatSpace}, the gauge field~\eqref{eq:KN_SUSY_GaugeField} is pure imaginary, which may seem bad. 
    It is important to recall, though, that the forms appearing in~\eqref{eq:KSW_Criterion_QuadraticForm} are fluctuations around the fixed background, whereas the curvature $\cF$ of~\eqref{eq:KN_SUSY_GaugeField} forms part of the background itself. 
    Therefore, there is no a priori contradiction with~\eqref{eq:KSW_Criterion_QuadraticForm}. A better understanding of the gauge field could come from the string theory embedding, but for ungauged supergravity this does not necessarily translate into a geometric question. \\
    On the other hand, one could hope to get a more refined control in top-down approaches to AAdS solutions, as one may argue that proper dual gravitational saddle points of the twisted and superconformal indices are not solutions to four/five-dimensional minimal gauged supergravity, but rather solutions of ten/eleven-dimensional string/M-theory---to which the KSW criterion should be applied. To give a concrete example, solutions $(Y_4,g,\cA)$ of the minimal gauged supergravity \eqref{eq:MinimalGaugedSUGRA_Action} can be uplifted to eleven dimensions on any seven-dimensional Sasaki--Einstein manifold $SE_7$ as~\cite{Gauntlett:2007ma}
    \begin{equation}
    \label{eq:Uplift_MinimalGauged}
    \begin{split}
        g(Y_{11}) &\=  \upliftL^2 \Biggl( \frac{1}{4} g(Y_4) + \ell^2 \biggl( \Bigl( \rd\psi + \sigma + \frac{1}{2\ell} \cA \Bigr)^2 + g(N_6) \biggr) \Biggr) \, , \\
        G_4 &\= \upliftL \Biggl( \frac{3}{8\ell} \, \vol(Y_4) - \frac{\ell^2}{2} *_4 \cF \wedge J \Biggr) \, .
    \end{split}
    \end{equation}
    Here $\partial_\psi$ is the Reeb vector field of the~$SE_7$, $J$ is the K\"ahler form on~$N_6$ 
    (the base of the~$U(1)$ fibration generating the $SE_7$) such that $\rd \sigma = 2J$, and $\upliftL > 0$ 
    is a constant that is fixed by the quantization of the four-form $G_4$ through the four-cycles of the $SE_7$. 
    For the saddles discussed in Section~\ref{sec:TTI}, the gauge field~\eqref{eq:TTI_Bulk_SUSY_Soln_GaugeField} is real in Euclidean signature, as are the eleven-dimensional metric and four-form, and thus the eleven-dimensional uplift is allowable.
    However, the situation is not as clean for the saddles dual to the superconformal indices in AAdS$_4$. 
    Uplifting the saddles of Section \ref{sec:AdS4} 
    using~\eqref{eq:Uplift_MinimalGauged} leads not only to a complex eleven-dimensional metric tensor (to which one could apply the KSW criterion), 
    but also to a complex four-form. 
    Therefore, the issue of the interpretation of the complex gauge field still persists even in the uplifted geometry.\\ 
    In order to address the allowability of such backgrounds, we need a criterion generalizing KSW to other background fields in string and M-theory.
\end{itemize}

\section*{Acknowledgements}

We are grateful to Davide Cassani, Jerome Gauntlett, Luca Iliesiu, Oliver Janssen, Don Marolf, Dario Martelli, Shiraz Minwalla, 
Rishi Mouland, Chintan Patel, Pietro Pelliconi, Mukund Rangamani, Luigi Tizzano, 
Gustavo Joaquin Turiaci, Edward Witten for helpful discussions. 
PBG is supported by the SNSF Ambizione grant PZ00P2$\_$208666.
S.M.~acknowledges the support of the J.~Robert Oppenheimer 
Visiting Professorship 2023-24 at the Institute for Advanced Study, Princeton, USA and 
the STFC grants ST/T000759/1,  ST/X000753/1 during the course of this work.
This work was performed in part at Aspen Center for Physics, 
which is supported by National Science Foundation grant PHY-2210452,
and in part  during the KITP program, “What is string theory? 
Weaving perspectives together”, which was supported by the grant NSF PHY-2309135 
to the Kavli Institute for Theoretical Physics (KITP).
We would like to thank the ACP, KITP, and UCSB Physics for hospitality.

\appendix

\section{Lorentzian solutions}
\label{app:Lorentzian_Solutions}

The solutions presented in the text can be obtained performing analytic continuations and imposing supersymmetry on various real Lorentzian black hole solutions. For completeness, in this appendix we collect them.

\medskip

We begin with four-dimensional Einstein--Maxwell theory 
\begin{equation}
    S = \frac{1}{16\pi} \int ( R - \cF^2) \vol \, , 
\end{equation}
from which by Wick rotation one obtains \eqref{eq:Action_Ungauged}. A solution of this theory is the Kerr--Newman black hole
\begin{align}
\label{eq:KN4_Metric}
\begin{split}
    \rd s^2 &= - \frac{\Delta_r}{B} \rd t^2 + W \left( \frac{\rd r^2}{\Delta_r} + \rd\BulkTheta^2 \right) \\
    & \ \ \ + \sin^2\BulkTheta \, B \left( \rd\phiCanonicalPeriodicities + \aL \frac{ \Delta_r \left( r_+^2 + \aL^2 \cos^2\BulkTheta \right) + (r^2 + \aL^2)(r^2 - r_+^2) }{(r_+^2 + \aL^2) B W} \rd t \right)^2 \, ,
\end{split} \\
\label{eq:KN4_GaugeField}
    \cA &= \frac{\qL r}{W} \left( (1 - \aL \sin^2\BulkTheta \, \Omega) \rd t - \aL \sin^2\BulkTheta \, \rd\phiCanonicalPeriodicities \right) -  \frac{\qL r_+}{r_+^2 + \aL^2} \rd t \, ,
\end{align}
where
\begin{equation}
\begin{split}
    \Delta_r &= r^2 + \aL^2 - 2 m r + \qL^2 \, , \qquad \qquad 
    W = r^2 + \aL^2 \cos^2\BulkTheta \, , \\ 
    B &= \frac{(r^2 + \aL^2)^2 - \aL^2 \sin^2\BulkTheta \, \Delta_r}{W} \, , \qquad 
    \Omega = \frac{\aL}{r_+^2 + \aL^2} \,,
\end{split}
\end{equation}
and $r_+$ is the largest solution to $\Delta_r=0$, namely $r_+ = m + \sqrt{m^2 - \aL^2-\qL^2}$. Here $r \geq r_+$, and $\theta \sim \theta + \pi$ and $\phiCanonicalPeriodicities\sim \phiCanonicalPeriodicities + 2\pi$ describe a $2$-sphere. This solution depends on the parameters $(m,a,q)$ and describes a black hole provided $m^2 \geq \aL^2 + \qL^2$. Let $V$ be the Killing generator of the black hole. We define the electric potential by
\begin{equation}
\label{eq:App_Phie}
    \Phi_e \equiv V^\mu \cA_\mu \rvert_{r=r_+} - V^\mu \cA_\mu \rvert_{r\to \infty} \, .
\end{equation}
For this black hole, $V=\partial_t$, and the electric potential is
\begin{equation}
    \Phi_e = \frac{\qL r_+}{r_+^2 + a^2} \, .
\end{equation}
The gauge is chosen such that $V^\mu \cA_\mu \rvert_{r=r_+} = 0$. In this family of black holes there is the extremal Reissner--Nordstr\"om black hole
\begin{align}
\label{eq:RN4_Metric}
\begin{split}
    \rd s^2 &= - \left( 1 - \frac{\qL}{r} \right)^2 \rd t^2 + \left( 1 - \frac{\qL}{r} \right)^{-2} \rd r^2 + r^2 \left( \rd\BulkTheta^2 + \sin^2\BulkTheta \, \rd\phiCanonicalPeriodicities^2 \right) \, ,
\end{split} \\
\label{eq:RN4_GaugeField}
    \cA &= \left( \frac{\qL}{r} - 1 \right) \rd t \, ,
\end{align}
which is a supersymmetric solution of minimal ungauged supergravity, as it supports a globally defined spinor solving \eqref{eq:KSE_Ungauged}, and depends on a unique parameter $q$. The solutions discussed in Section~\ref{sec:FlatSpace} are supersymmetric non-extremal deformations of this solution.

\medskip

We move to four-dimensional Einstein--Maxwell theory with a negative cosmological constant
\begin{equation}
\label{eq:MinimalGaugedSUGRA_Action_Lorentzian}
    S = \frac{1}{16\pi} \int \left( R + \frac{6}{\ell^2} - \cF^2 \right) \vol \, , 
\end{equation}
from which by Wick rotation one obtains \eqref{eq:MinimalGaugedSUGRA_Action}. This theory admits an AdS$_4$ solution with radius $\ell$.
The first family of solutions that we are interested in are the static dyonic black holes with a horizon given by a Riemann surface $\Sigma_g$ with genus $g>1$
\begin{align}
\begin{aligned}
    \rd s^2 &= - V(r) \, \rd t^2 + \frac{\rd r^2}{V(r)} + r^2 \left( \rd \BulkTheta^2 + \sinh^2\BulkTheta \, \rd\phi^2 \right) \, , \\
    V(r) &= -1 + \frac{r^2}{\ell^2} - \frac{2\eta}{r} + \frac{q^2 + p^2}{r^2} \, , \\[5pt]
    \cA &= \frac{\qL}{r} \rd t + p \cosh\theta \, \rd\phi - \frac{\qL}{r_h} \rd t\, , 
\end{aligned}
\end{align}
where $r\geq r_h$, the largest positive root of $V(r)$, and we have chosen a metric of constant curvature on $\Sigma_g$ obtained by taking a quotient of $H^2$ (parametrized by $\theta$ and $\phi$) and normalizing so that $\vol(\Sigma_g) = 4\pi (g-1)$.
This metric depends on three parameters $(\eta, p, q)$ and describes a black hole provided \cite{Caldarelli:1998hg}
\begin{equation}
\label{eq:AdSTopological_Condition}
    \eta \geq \eta_0(q,p) \equiv \frac{\ell}{3\sqrt{6}} \left( \sqrt{1 + 12 \frac{q^2 + p^2}{\ell^2}} - 2 \right) \sqrt{ \sqrt{1 + 12 \frac{q^2 + p^2}{\ell^2}} +1 } \, ,
\end{equation}
whereas for $\eta < \eta_0(q,p)$ it's a naked singularity. The eletric potential for this solution is
\begin{equation}
    \Phi_e = \frac{q}{r_h} \, .
\end{equation}
The action \eqref{eq:MinimalGaugedSUGRA_Action_Lorentzian} is the bosonic action of the minimal gauged supergravity with Killing spinor equation \eqref{eq:MinimalGaugedSUGRA_KSE}, and among the black holes above sits the supersymmetric extremal static magnetically charged black hole with Riemann surface horizon \cite{Caldarelli:1998hg}
\begin{align}
\begin{aligned}
    \rd s^2 &= - \left( \frac{r}{\ell} - \frac{\ell}{2r} \right)^2 \, \rd t^2 + \left( \frac{r}{\ell} - \frac{\ell}{2r} \right)^{-2} \rd r^2 + r^2 \left( \rd \BulkTheta^2 + \sinh^2\BulkTheta \, \rd\phi^2 \right) \, , \\
    \cA &= \pm \frac{\ell}{2} \, \cosh\theta \, \rd\phi \, .
\end{aligned}
\end{align}
Note that this solution doesn't have any parameter, beside the genus of the Riemann surface. Supersymmetric deformations of this black hole have been discussed in Section~\ref{sec:TTI}.

\medskip

The second family of solutions is the AdS-Kerr--Newman black holes
\begin{align}
\label{eq:Metric_Four_Dimensional_Minimal}
\begin{split}
\rd s^2 &= - \frac{\Delta_r\Delta_\BulkTheta}{B\Xi^2}  \rd t^2 + W \left( \frac{\rd r^2}{\Delta_r} + \frac{\rd \BulkTheta^2}{\Delta_\BulkTheta}  \right) \\
    & \ \ \ + \sin^2\BulkTheta \, B \left( \rd \phiCanonicalPeriodicities + a \frac{ \Delta_r \left( r_+^2 + a^2 \cos^2\BulkTheta \right) + \Delta_\BulkTheta ( r^2 + a^2) (r^2 - r_+^2) }{ (r_+^2 + a^2) B W \Xi} \rd t\right)^2   \, , 
\end{split}\\[10pt]
    \cA &= \frac{ m r \sinh \delta}{W\Xi}\Big( \left( \Delta_\BulkTheta - a \sin^2\BulkTheta \, \Omega \right) \rd t - a \sin^2\BulkTheta \, \rd \phiCanonicalPeriodicities \Big) - \frac{ m r_+ \sinh \delta }{a^2 + r_+^2} \, \rd t  \, .
\end{align}
Here $r \geq r_+$ is the largest positive root of $\Delta_r$, and $\BulkTheta\sim \BulkTheta + \pi$, $\phiCanonicalPeriodicities\sim \phiCanonicalPeriodicities + 2\pi$, and
\begin{equation}
\label{eq:Quantites_Four_Dimensional_Minimal_Rotating}
\begin{split}
    & \Delta_r = (r^2 + a^2)(1 + r^2/\ell^2) - 2 m r \cosh \delta + m^2 \sinh^2 \delta \, , \\
    & \Delta_\BulkTheta = 1 -  a^2 /\ell^2 \, \cos^2 \BulkTheta \, , \qquad W = r^2 + a^2 \cos^2 \BulkTheta \, , \qquad \Xi = 1 - a^2 /\ell^2 \, , \\
    & B \equiv \frac{\Delta_\BulkTheta(r^2+a^2)^2 - a^2 \sin^2\BulkTheta \, \Delta_r}{W \Xi^2} \, , \\
    & \Omega = a\frac{1 + r_+^2/\ell^2}{a^2 + r_+^2} \, , \\
    & \Phi_e = \frac{ m r_+ \sinh \delta }{a^2 + r_+^2} \, .
\end{split}
\end{equation}
This family of black holes is described by three parameters $(m,a,\delta)$.
In this family sits the supersymmetric extremal rotating electrically charged black hole
\begin{align}
\label{eq:Metric_Four_Dimensional_Minimal_Rotating_BPS}
\begin{split}
\rd s^2 &= - \frac{\Delta_r\Delta_\BulkTheta}{B\Xi^2}  \rd t^2 + W \left( \frac{\rd r^2}{\Delta_r} + \frac{\rd \BulkTheta^2}{\Delta_\BulkTheta}  \right) \\
    & \ \ \ + \sin^2\BulkTheta \, B \left( \rd \phiCanonicalPeriodicities + \frac{ r_\star^2 \Delta_r \left( 1 + r_\star^2 /\ell^2 \cos^2\BulkTheta \right) + \Delta_\BulkTheta ( r^2 + r_\star^4 /\ell^2) (r^2 - r_\star^2) }{  \ell (1 + r_\star^2 /\ell^2) B W \Xi} \rd t\right)^2   \, ,
\end{split}\\[10pt]
    \cA &= \frac{ r_\star }{W (1-r_\star^2/\ell^2) }\left[ \left( \Delta_\BulkTheta - r_\star^2/\ell^2 \sin^2\BulkTheta \right) \rd t - r_\star^2/\ell \sin^2\BulkTheta \, \rd \phiCanonicalPeriodicities \right] - \rd t  \, , \\[10pt]
\begin{split}
    & \Delta_r = \left( r - r_\star \right)^2 \left( r_\star^4/\ell^4 + (r^2 + 2r r_\star + 3r_\star^2)/\ell^2 + 1 \right) \, , \\
    & \Delta_\BulkTheta = 1 -  r_\star^4 /\ell^4 \, \cos^2 \BulkTheta \, , \qquad W = r^2 + r_\star^4/^2 \cos^2 \BulkTheta  \, , \qquad \Xi = 1 - r_\star^4 /^4 \, , \\
    & B \equiv \frac{\Delta_\BulkTheta(r^2+r_\star^4 /^2)^2 - r_\star^4/^2 \sin^2\BulkTheta \, \Delta_r}{W \Xi^2} \, .
\end{split}
\end{align}
This depends on a single parameter $r_\star$, which is the location of the horizon.
Supersymmetric deformations of this black hole have been discussed in Section~\ref{sec:AdS4}.

\bibliographystyle{JHEP}
{\small
\bibliography{Bib_BH}}

\end{document}